%% file: main.tex
\documentclass[aps,prd,preprintnumbers,twocolumn,nofootinbib]{revtex4-2}

\usepackage{graphicx}
\usepackage{amsmath,amssymb}
\usepackage{float}
\usepackage{xcolor}
\usepackage{multirow}
\usepackage{bm}
\usepackage{siunitx}

\usepackage{hyperref}
\hypersetup{
    colorlinks=true,     
    linkcolor=blue,      
    citecolor=blue,      
    filecolor=blue,      
    urlcolor=blue        
}

\usepackage[nameinlink]{cleveref}

\newcommand{\NLPM}{$\text{NLO}+\text{LPM}^{\text{LO}}$}

\newcommand{\tinytext}[1]{\text{\tiny{#1}}}

  \def\gsim{\mathrel{\rlap{\lower0.25em\hbox{$\sim$}}\raise0.2em\hbox{$>$}}} 
  \def\lsim{\mathrel{\rlap{\lower0.25em\hbox{$\sim$}}\raise0.2em\hbox{$<$}}}
  \def\lg{\mathrel{\rlap{\lower0.25em\hbox{$>$}}\raise0.25em\hbox{$<$}}}
  \def\gl{\mathrel{\rlap{\lower0.25em\hbox{$<$}}\raise0.25em\hbox{$>$}}}

\begin{document}

\title{Lattice QCD estimates of thermal photon production from the QGP} 

\author{Sajid Ali$^{a,f}$, Dibyendu Bala$^a$, Anthony Francis$^b$, Greg Jackson$^{c,d}$, Olaf Kaczmarek$^a$, Jonas Turnwald$^e$, Tristan Ueding$^a$, Nicolas Wink$^e$
}

\affiliation{
$^a$Fakult\"at f\"ur Physik, Universit\"at Bielefeld, D-33615 Bielefeld, Germany\\
$^b$Institute of Physics, National Yang-Ming Chiao Tung University, 30010 Hsinchu, Taiwan\\
$^c$Institute for Nuclear Theory, Box 351550, University of Washington, Seattle, WA 98195-1550, United States\\
$^d$SUBATECH (Nantes Universit\'e, IMT Atlantique, IN2P3/CNRS), 4 rue Alfred Kastler, La Chantrerie BP 20722, 44307 Nantes, France\\
$^e$Institut f\"ur Kernphysik, Technische Universit\"at Darmstadt, D-64289 Darmstadt, Germany\\
$^f$Government College University Lahore, Department of Physics, Lahore
54000, Pakistan
}

\collaboration{HotQCD Collaboration}

\date{\today}

\begin{abstract}
Thermal photons produced in heavy-ion collision experiments are an important observable for understanding quark-gluon plasma (QGP). The thermal photon rate from the QGP at a given temperature can be calculated from the spectral function of the vector current correlator. Extraction of the spectral function from the lattice correlator is known to be an ill-conditioned problem, as there is no unique solution for a spectral function for a given lattice correlator with statistical errors. The vector current correlator, on the other hand, receives a large ultraviolet contribution from the vacuum, which makes the extraction of the thermal photon rate difficult from this channel. We therefore consider the difference between the transverse and longitudinal part of the spectral function, only capturing the thermal contribution to the current correlator, simplifying the reconstruction significantly. The lattice correlator is calculated for light quarks in quenched QCD at $T=470~$MeV ($\sim 1.5\, T_c$), as well as in 2+1 flavor QCD at $T=220~$MeV ($\sim 1.2 \, T_{pc}$) with $m_{\pi}=320$ MeV. In order to quantify the non-perturbative effects, the lattice correlator is compared with the corresponding \NLPM\ estimate of correlator. The reconstruction of the spectral function is performed in several different frameworks, ranging from physics-informed models of the spectral function to more general models in the Backus-Gilbert method and Gaussian Process regression. We find that the resulting photon rates agree within errors.
\end{abstract}

\maketitle
\section{Introduction}
\label{sec:intro}
\input{Intro.tex}

\section{Lattice details}
\label{sec:lattice_details}
\input{lattice_details.tex}

\section{Comparison with perturbative estimate}
\label{sec:comp_pt}
\input{comp_with_pt.tex}

\section{Spectral reconstruction}
\label{sec:spec}
\input{spectral_reconstruction.tex}
\section{\texorpdfstring{Comparison of $D_{\text{eff}}T$}{Comparison of DeffT}}
\label{sec:deff}
\input{Deff.tex}

\section{Conclusion}
\label{sec:conclusion}
\input{conclusion.tex}

\appendix
\input{app.tex}

\newpage
\bibliography{ref}

\end{document}

%% file: Intro.tex
One of the important predictions of quantum chromodynamics (QCD), the theory of strong interactions, is that the QCD coupling constant decreases with the energy scale. This characteristic makes QCD strongly coupled (non-perturbative) at low energies, and weakly coupled at very high energies. As a result, non-perturbative studies of QCD are important at low energies. First-principles lattice QCD equation-of-state calculations show that, as the temperature increases, QCD matter undergoes a crossover transition from a hadronic phase to a quark-gluon plasma phase (QGP) at a pseudo-critical temperature of $T_{pc}={156.5\pm1.5}~\textrm{MeV}$~\cite{HotQCD:2018pds,HotQCD:2014kol,Borsanyi:2013bia}. 

Experimentally, this phase has been studied in ultrarelativistic heavy-ion collisions at RHIC and LHC. Photons and dileptons produced in a heavy-ion collision emerge directly from the point of creation without any further interaction with their surroundings~\cite{Pisarski1981,McLerran1984,Kajantie:1981wg,Kajantie:1986dh,Gale1987}. As a result, they carry information about the local region around their place of origin. However, in experiment, one measures the total number of photons produced from different stages of the QCD matter evolution after the collision. Separating photons produced at different stages of evolution is a challenging task, as the vast majority ($\sim 80...90$\%) of photons emitted because of hadronic decays (e.g.  $\pi^0 \to \gamma \,\gamma$) during the hadronization stage \cite{Turbide:2003si}. 
These photons are known as the hadronic background. The remaining photons, which are produced directly from the QGP or before its formation, are called {\em direct} photons. Extraction of these direct photons from the hadronic background requires sophisticated techniques and careful analysis~\cite{PHENIX:2014nkk}. 

The Au-Au collisions at PHENIX~\cite{PHENIX:2008uif} and Pb-Pb collisions at ALICE~\cite{Wilde:2012wc} have revealed large yields of direct photons. The high-$p_T$ segment of the photon spectra (prompt photons) can be related to the yield of photons generated in proton-proton collisions, scaled according to the number of participating binary collisions. The excess photons observed in the low-$p_T$ region of the spectra may be tentatively attributed to in-medium plasma effects. Measurements of direct photons and dileptons also reveal a large elliptical flow, quantified by the coefficients $v_2$~\cite{PHENIX:2015igl,Lohner_2013,Chatterjee:2005de}. However, understanding the origin of this phenomenon from first principles poses a significant theoretical challenge~\cite{PhysRevC.84.054906}.

Any attempt to explain these observables must involve integrating the (multistage) hydrodynamic expansion of the plasma with the differential photon or dilepton production rate embedded at each stage of its evolution. 
Simulations have revealed that the `slopes' of the $p_T$ spectrum (for photons)
and the invariant mass spectrum (for dileptons) can be systematically related to an average temperature of the fireball~\cite{Shen:2013vja,Churchill:2023vpt,Churchill:2023zkk}. 
The anisotropic flow of direct photons and dileptons has been shown to be sensitive to the initial conditions and shear viscosity~\cite{Chaudhuri:2011up,PhysRevC.93.044906,Vujanovic:2016anq}. Since electromagnetic probes are generated at every stage of plasma evolution, they may supply useful constraints on the viscosity and other transport coefficients. One of the important ingredients in such analyses is a precise understanding of the photon production mechanism from each stage of the plasma evolution. Since hydrodynamic expansion assumes local thermal equilibrium, computation of the photon production rate at each stage requires the use of equilibrium thermal field theory calculations.

Numerous studies have investigated the thermal photon rate, both perturbatively and non-perturbatively, using lattice techniques. 
The perturbative evaluation of the rate has a long history, and was
regarded as settled in the early 1990s when two groups independently arrived at the same answer~\cite{Baier:1991em,Kapusta:1991qp}. 
Those works implemented resummed thermal fermion propagators to account for screening, the first step needed to obtain a finite result. 
It was then discovered that those treatments were incomplete and additional
processes (naively of higher order) were incorporated to establish the full ${\cal O}(\alpha_{\rm em}\alpha_s)$ result~\cite{Arnold:2001ba,Arnold:2001ms}.
More recently, it was understood how to take advantage of a Euclideanization property for thermal light-cone correlators to attain relative ${\cal O}(\sqrt{\alpha_s})$ corrections~\cite{Ghiglieri:2013gia}.

On the lattice QCD side, a major challenge is the fact that the thermal production of photons requires real-time information. 
This can only be achieved after analytic continuation of Euclidean data, which is formally an ill-conditioned inverse operation~\cite{Cuniberti:2001hm, Meyer:2011gj}. Consequently, various techniques and models have been employed for spectral reconstruction. In Ref.~\cite{Ghiglieri:2016tvj}, the photon rate was estimated using the vector channel correlator, wherein the UV part of the spectral function was calculated perturbatively, while the IR part was fixed using lattice data. A similar method was employed in another study~\cite{Ce:2022fot}, where the photon rate was estimated from the transverse channel correlator. A new correlator was utilized in Ref.~\cite{Ce:2020tmx}, based on the difference between the transverse and longitudinal channels. This approach is advantageous because it suppresses the UV part of the spectral function. Recently, a novel idea has been proposed for calculating the photon rate. In this method, the photon production rate is studied using the imaginary momentum correlator, which does not require spectral reconstruction~\cite{Meyer:2018xpt,Ce:2023oak}.

In this paper, following ~\cite{Ce:2020tmx}, we estimate the thermal photon rate from the QGP using the difference between the transverse and longitudinal correlators. For the spectral reconstruction, we used various techniques, ranging from physics-informed model fits and the Backus-Gilbert method to Gaussian process regression.

Using these methods, we can calculate the thermal photon rate from the lattice correlator for light quarks in two different scenarios: Quenched QCD at a temperature of $\SI{470}{\mega\eV}$ ($1.5\,T_c$), and (2+1)-flavor QCD at a temperature of $\SI{220}{\mega\eV}$ ($1.22\,T_{pc}$). For the quenched QCD scenario, we have extrapolated the correlator to the continuum, which means that our prediction for the photon production rate is also at the continuum. For the full QCD scenario, our results are at a finite lattice spacing. 

For the full QCD case, we use {\em highly improved staggered quark} (HISQ) configurations for the background gauge fields. This is advantageous because generating HISQ configurations is easier than generating dynamical Wilson fermion configurations, especially at the fine lattice spacing we are considering. One reason for this is that staggered fermions preserve a remnant of chiral symmetry, which protects the Dirac operator from having small real eigenvalues responsible for exceptional configurations. Additionally, chiral symmetry prevents the quark mass from having any additive renormalization constant, making quark mass tuning much easier. In the valence sector, we use clover-improved Wilson fermions. The reason for this choice is that HISQ correlators contain both oscillating and non-oscillating modes, which complicates the spectral reconstruction method at finite temperatures. The downside of this approach is the presence of mixed-action cut-off effects. However, the mixed-action effect, quantified in terms of the relative mass difference ~\cite{Zhao:2022ooq}, turns out to be small at the lattice spacing we are working with. We therefore expect these mixed-action effects to vanish in the continuum limit.

The paper is organized into the following sections: In \Cref{sec:t-l_spectral}, we will briefly discuss the thermal photon rate and the correlator that we used to calculate it. In  \Cref{sec:lattice_details}, we present the lattice details and perform the continuum extrapolation of quenched lattice data. In \Cref{sec:comp_pt}, we compare the lattice correlator with the correlator obtained from perturbative spectral function, to quantify the non-perturbative effects. We undertake the spectral reconstruction using multiple techniques in \Cref{sec:spec}. Finally, in \Cref{sec:conclusion}, we compare the photon production rates obtained from different techniques and provide the final value for the photon production rate.

\section{Thermal photon rate and Spectral function}
\label{sec:t-l_spectral}

In a thermalized plasma at temperature $T$, the photon production rate is defined as the number of photons radiated from the plasma per unit time and per unit volume. 
We consider the thermal photon production rate  due to $N_f=3$ degenerated quark flavors, 
whose total electromagnetic current that couples to the photon is given by 
$e \sum_{f} Q_{f} \bar \psi_{f} \gamma^{\mu} \psi_{f}$\,, 
where $Q_f$ is the quark charge fraction 
and $\psi_{f}$ is the quark field of flavor $f$. 
If we consider the case $N_f = 3$  where $f = {u,d,s}$, then
$\sum_{f} Q_f=0$ and $\sum_{f} Q_{f}^2=\frac{2}{3}$. Under these conditions, the disconnected\footnote{In terms of the $\alpha_s$-expansion, the disconnected contribution only starts at ${\cal O}(\alpha_s^3)$~\cite{Baikov:2012zn}.} quark contribution vanishes exactly and therefore to leading order in the QED coupling $\alpha_{\rm em}$, this rate can be calculated as
\begin{equation}\label{eq:photonprod}
    \frac{{\rm d}\Gamma_\gamma}{{\rm d}^3{\vec k}}
    \;=\;
    -\, \frac{\alpha_{\rm em} n_b(k)}{2 \pi^2 k}
    \Big\{ {\textstyle\sum_{i=1}^{N_f} Q_i^2} \Big\}
    g^{\mu\nu}\rho_{\mu\nu}\big(\,\omega=|\vec k|,\,\vec k\,\big)
    \;,
\end{equation}
where 
$n_b(\omega) = 1/(e^{\omega/T}-1)$ is the Bose distribution function~\cite{McLerran:1984ay}. 
We take the Minkowski metric as $g_{\mu\nu} = {\rm diag}(1,-1,-1,-1)$. In this expression, the $\rho_{\mu\nu}$ is the connected part of the vector current spectral function for a single flavor.

 The spectral function $\rho_{\mu\nu}(\omega,\vec k)$ contains the full information about QCD dynamics and can be calculated through analytic continuation from the Matsubara frequency modes,\footnote{These discrete energies are denoted $\omega_n = 2\pi T n\,$, where $n\in \mathbb{Z}$.} i.e., $\rho_{\mu\nu}(\omega, \vec k)=\mathrm{Im}[G^{E}_{\mu\nu}(\omega_n\to -i (\omega+i0^+), \vec k)]$, where the Euclidean current correlator is given by
\begin{equation}\label{eq:KL_definition}
    G^E_{\mu\nu}(\omega_{n},\vec k) = \int {\rm d}^3\vec x \int_{0}^{\beta} {\rm d}\tau\, e^{i(\omega_{n} \tau-\vec k.\vec x)}\langle J_{\mu}(\tau, \vec x) J_{\nu}(0, \vec 0)\rangle
    \,,
\end{equation}
where $\beta = 1/T$ is the temporal extent and the electromagnetic current $J_{\mu}(\tau,\vec x)=\bar \psi(\tau,\vec x)\gamma_{\mu} \psi(\tau,\vec x)\,$.

On the lattice we calculate the Euclidean correlation function $G^{E}_{\mu\nu}(\tau,\vec k)=T\,\sum _{n} \exp(i\omega_n \tau) G^E_{\mu\nu}(\omega_{n},\vec k)$. This correlator can be written in terms of the spectral function as
\begin{equation}\label{eq:KL}
	G^{E}_{\mu\nu}(\tau, \vec k)=\int_{0}^{\infty} \frac{{\rm d}\omega}{\pi} \rho_{\mu\nu}(\omega,\vec k) \frac{\cosh[\omega(\tau-\frac{1}{2T})]}{\sinh[\frac{\omega}{2T}]}
	\,.
\end{equation}

As mentioned before, extracting the spectral function from the Euclidean correlator is an ill-conditioned problem. Therefore, when inverting \Cref{eq:KL} naively, any errors in the lattice data lead to exponentially large errors in the corresponding spectral function. There are, consequently, many possible spectral functions that could reproduce the lattice data. The incorporation of additional, physics motivated, information is needed in order to `regularize' the problem and achieve reasonable reconstructions. 

The contracted 
(or ``vector channel'') spectral function $g^{\mu\nu} \rho_{\mu\nu}$,
appearing in \Cref{eq:photonprod},
is proportional to $\omega^2$ in the ultraviolet (UV) domain.
As a result, the lattice correlator, determined
by the integration in \Cref{eq:KL}, 
 receives a significant contribution from the UV part of the spectral function. 
 This fact complicates the reconstruction process in general, 
 as the thermal photon rate is dominated by the infrared (IR) part of the spectral function.
 Therefore, following Ref.~\cite{Ce:2020tmx}, we will estimate the thermal photon rate from the difference between the transverse and longitudinal (T-L) spectral functions. 
 This approach has the advantage in that the UV part of the spectral function is heavily suppressed, meaning that the corresponding Euclidean correlator is governed by the IR part of the spectral function, resulting in more reliable reconstructions. 

In order to obtain the T-L part, the spectral function $\rho_{\mu\nu}$ is decomposed in terms of $\rho_{T}$ (the transverse component) and $\rho_{L}$ (the longitudinal component) of the spectral function as
\begin{equation}
    \rho_{\mu\nu}(\omega,\vec k)=P^{T}_{\mu\nu} \rho_{T}(\omega,\vec k) +P^{L}_{\mu\nu} \rho_{L}(\omega,\vec k)
    \,.
\end{equation}
Here, $P^{T}_{\mu\nu}$ and $P^{L}_{\mu\nu}$ are the projection operators, given explicitly by
\begin{equation}
\begin{split}
    & P^{T}_{ij}(\omega,\vec k) \;= \; -g_{ij}-\frac{k_{i}k_{j}}{\vec k^2}\,,\\[1ex]
    & P^{T}_{0i} \;=\; P^{T}_{i0} \;=\; P^{T}_{00} \;=\; 0\,,\\[1ex]
    & P^{L}_{\mu\nu}(\omega,\vec k) \;=\; -g_{\mu\nu}+\frac{K_{\mu}K_{\nu}}{K^2}-P^{T}_{\mu\nu}(\omega,\vec k)
    \,,
\end{split}
\end{equation}
where $K=(\omega,\vec k)$ is the associated four-momentum.
The projection operators satisfy the following relations: $P_T^2=P_T$, $P_L^2=P_L$, and $P_T P_L=P_L P_T=0\,$. In terms of these components, the vector channel spectral function is given by $-\rho^{\mu}_{\mu}=2\rho_T+\rho_L$. The photon production rate is therefore proportional to $\rho_{T}(\omega=|\vec k|,\vec k)$, as on the light cone longitudinal part of the spectral function vanishes, $\rho_L(\omega=|\vec k|,\vec k)=0\,$. This allows us to calculate the photon production rate using the T-L spectral function as
\begin{equation}
    \rho_H(\omega, \vec k) 
    \; = \; 
    2\big\{\,\rho_{T}(\omega,\vec k)-\rho_{L}(\omega,\vec k)\,\big\}
    \,.
\end{equation}
At zero temperature, because of the Ward identity and Lorentz invariance, the T-L spectral function vanishes, $\rho_H(\omega,\vec k)=0$. Therefore, non-zero $\rho_{H}(\omega, \vec k)$ displays purely thermal effects. Using the Operator Product Expansion (OPE), it has also been shown that in the domain where $\omega \gg k,\pi\,T $, this spectral function behaves asymptotically like $1/\omega^4$~\cite{Caron-Huot:2009ypo, Brandt:2017vgl}. One consequence of the OPE, is that this spectral function satisfies a sum rule,
\begin{equation} \label{eq:sumrule}
    \int_{0}^{\infty}\, {\rm d}\omega \, \omega \,\rho_{H}(\omega)=0
    \,.
\end{equation}
We should note that the photon production rate \Cref{eq:photonprod} is often (equivalently) expressed 
as
\begin{equation}
    \frac{{\rm d}\Gamma_{\gamma}}{{\rm d}^3{\vec k}}
    \;=\;
    \frac{\alpha_{\rm em} n_{b}(k)\chi_{q}}{ \pi^2 } 
    \Big\{ {\textstyle\sum_{i=1}^{N_f} Q_i^2} \Big\}
    D_{\text{eff}}(k)
    \,,
\end{equation}
where the effective diffusion coefficient is 
defined by
\begin{equation}\label{eq:Deff}
D_{\text{eff}}(k) \; \equiv \; \frac{\rho_H\big(|\vec k|,k\big)}{2 \chi_q |\vec k|}
\,.
\end{equation}
Here $\chi_q$ is the quark number susceptibility (defined later in \Cref{chi_q}). 
In the hydrodynamic regime of large wavelengths, $D_{\text{eff}}(k)$ approaches the well-known diffusion coefficient $D$, i.e.~$\lim_{k \to 0} D_{\text{eff}}(k)=D$.

%% file: lattice_details.tex
In lattice QCD calculations, space-time has finite spatial extent ($L$) and discrete momenta, given by $\vec k = \left( \frac{2\pi n_x}{L},\frac{2\pi n_y}{L},\frac{2\pi n_z}{L}\right)$, where $(n_x,n_y,n_z) \in \mathbb{Z}$. Correlation functions $G_{xx}(\tau, \vec k)$, $G_{yy}(\tau, \vec k)$, $G_{zz}(\tau, \vec k)$, and $G_{\tau \tau}(\tau, \vec k)$ are computed on the lattice. If the momentum is in the the $x$ direction i.e. $\vec k=(k, 0, 0) $, then the correlator for $\rho_H$ is given by \cite{Brandt:2017vgl}:
\begin{widetext}
\begin{equation}
G_{H}(\tau,k_x=k) \ =\ 
G_{yy}(\tau,k_x=k)\,+\,G_{zz}(\tau,k_x=k)\,-\,
2\big(\, G_{xx}(\tau,k_x=k)-G_{\tau \tau}(\tau,k_x=k) \,\big)\,.
 \end{equation}
The lattice correlator is symmetric under $\tau\to \beta-\tau$ and due to rotational invariance, $G_{H}(\tau,k_x=k)=G_{H}(\tau,k_y=k)=G_{H}(\tau,k_z=k)$. 
Therefore, we considered the following quantity as an estimate of the T-L correlator:
\begin{equation}
 G_{H}(\tau, k)\ =\ 
 \frac16 
 \Big\{ \,
 G_{H}(\tau,k_x=k)+G_{H}(\tau,k_y=k)+G_{H}(\tau,k_z=k)+(\tau\to\beta-\tau)
 \,\Big\}\,.
 \end{equation}
\end{widetext}
\begin{table}
\centering
\begin{tabular}{c@{\hspace{0.5em}}c@{\hspace{1em}}c@{\hspace{1em}}c@{\hspace{1em}}c@{\hspace{1em}}c@{\hspace{1em}}c}
\hline \\[-1.5mm]
$N_f$ & $T$ & $\beta_0$ & $\kappa$ &$N_\sigma^3\times N_\tau$ & confs & $a^{-1}$[GeV]  \\[1.5mm] 
\hline
 \\[-2.5mm] 
\multirow{3}{*}{$0$} & \multirow{3}{*}{$1.5\,T_c$} & $7.394$ & $0.13407$ & $120^3 \times 30$ & $1950$ & $14.1$ \\[0.5mm] 
& & $7.192$ & $0.13440$ & $\ 96^3\times 24$ & $2000$ & $11.3$ \\[0.5mm]  
& & $7.035$ & $0.13467$ & $\ 80^3 \times 20$ & $1824$ & $9.39$ \\[0.5mm] 
\hline  
\\[-2.5mm]
$2+1$ & $1.2\,T_{pc}$ & $8.249$ & $0.13515$ & $\ 96^3 \times 32$ & $1750$ & $7.04$ \\[0.5mm]  \hline
\end{tabular}
\caption{
Details of the lattices studied in this paper, where $\beta_0$ is the coefficient of the Wilson plaquette and $\kappa$ is the hopping parameter. The  critical temperature  $T_c=\SI{313}{\mega\eV}$ for $N_f=0$ case and the pseudo critical temperature is $T_{pc}=\SI{180}{\mega\eV}$ for $N_f=2+1$ case.} 
\label{tab:lat-detail}
\end{table}
We calculated the $G_H$ correlator using Clover-improved Wilson fermions for quenched configurations at a temperature of $\sim1.5 \,T_c$ and (2+1)-flavor QCD configurations at a temperature of $\sim 1.2 \,T_{pc}$. The scale for quenched QCD has been set by $r_0=\SI{0.47}{\femto\meter}$~\cite{Sommer:2014mea} and the $T_c$ is obtained by $T_c=0.7457/r_0$ \cite{Francis:2015lha}. In the case of full QCD, the scale is determined by $r_1=\SI{0.3106}{\femto\meter}$ and $T_{pc}=\SI{180}{\mega\eV} $ was obtained from disconnected chiral susceptibility, cf.~Ref.~\cite{MILC:2010hzw}, as in Ref.~\cite{Altenkort:2023eav}.  

We generated the quenched configurations using a standard Wilson gauge action with heat-bath and overrelaxation updates. Each configuration was separated by 500 sweeps, with each sweep corresponding to one heat bath followed by four overrelaxation steps. We generated pure gauge configurations in this way for three lattice spacings, which allows us to perform a continuum extrapolation of the correlators. For the valence sector, we selected a hopping parameter $\kappa$ value very close to the $\kappa_c$ value, which was obtained by cubic spline interpolation of $\kappa_c$ values given in Ref.~\cite{Luscher:1996ug}. The clover improvement constant, also known as the Sheikholeslami–Wohlert coefficient ($c_{SW}$), was estimated from the non-perturbative parametrization of Ref.~\cite{Luscher:1996ug}.

For full QCD, we used gauge field configurations generated by the HotQCD collaboration 
using Highly Improved Staggered Quark (HISQ)~\cite{Follana:2006rc} and a tree-level improved L\"uscher-Weisz gauge action~\cite{Luscher:1984xn, LUSCHER1985250}. We used the RHMC algorithm~\cite{Clark:2003na} for the production of configurations and every fifth configuration has been saved for measurement. These configurations have also been used in ~\cite{Altenkort:2023eav}, and the auto-correlation time is estimated to be 10--30 configurations from a gradient flowed Polyakov loop. We have also looked at the gradient flowed topological charge, which is known to exhibit a larger auto-correlation time. We find that our gauge configurations are mostly in the trivial topological sector. This is expected, as at high temperatures, the topological susceptibility is small.  However, the correlation functions we consider here are not too sensitive to topology, and therefore, we expect the correlator to exhibit only a mild dependence on the topological charge. This behavior has been observed in Ref.~\cite{Ce:2020tmx}. We tuned the $\kappa$ value to match the pion mass, which in this study was unphysical and equal to $\SI{320}{\mega\eV}$ with $m_l=m_s/5$. The $c_{SW}$ coefficient was taken to be the tadpole-improved tree-level value, i.e.~$c_{SW}=1/u_0^3$, where $u_0$ is the fourth root of the plaquette expectation value. In all cases considered, the quark mass was significantly smaller than the temperature being studied. More detailed information regarding the lattice parameters can be found in \Cref{tab:lat-detail}.
\begin{figure*}
    \centering
    \includegraphics[width=8cm]{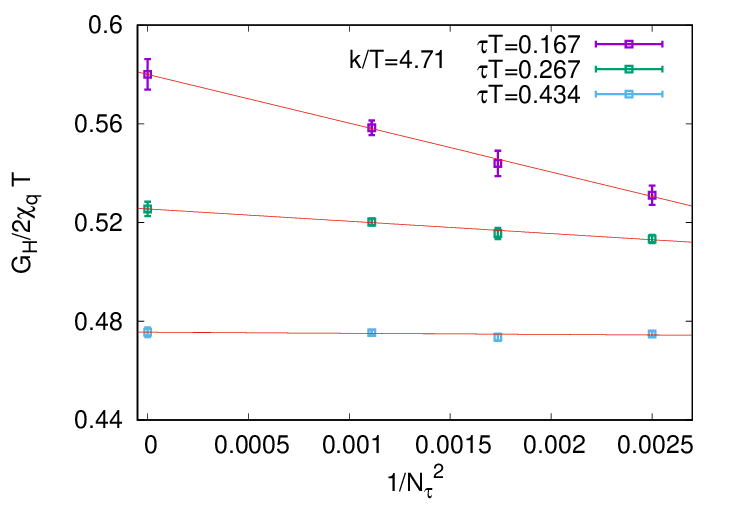}
    \includegraphics[width=8cm]{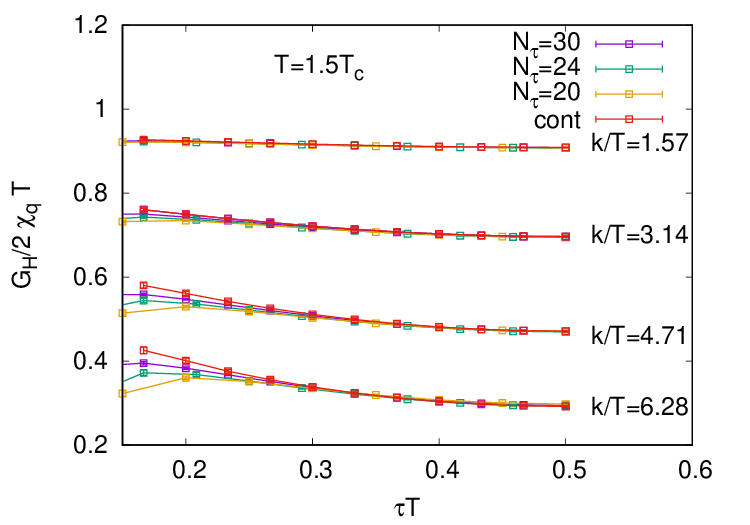}
	\caption{ Euclidean correlator from lattice simulations in the T-L channel,  cf.~\Cref{GH lattice}, for quenched QCD. Continuum extrapolation for various values of $\tau T$ at a fixed temperature appear on the left, and a comparison of the continuum correlator with the correlator calculated on finite lattice spacing for various spatial momenta is on the right.}
    \label{cont_corr}
\end{figure*}
Our implementation of the current operator on the lattice is not $O(a)$ improved, but as discussed in Ref.~\cite{Brandt:2017vgl}, the effect of this improvement is quark mass suppressed in the chirally symmetric phase. For the quenched case, the effect of the improvement coefficient is small in the coupling range we are working on~\cite{Guagnelli:1997db}. Therefore, we expect the correction to the continuum result should be $O(a^2)$ in both quenched and full QCD lattice correlator.

The correlator $G_H$ calculated on the lattice requires renormalization prior to performing the continuum limit. This correlator is multiplicatively renormalizable, which allows us to extrapolate to the continuum by considering the quantity $G_H/(2\chi_q T)$. Here, $\chi_q$ denotes the quark number susceptibility and in the continuum is defined as
\begin{equation}
\chi_q=\int_0^{1/T}\,{\rm d}\tau\, G_{\tau\tau}(\tau,\vec 0) \,.
\label{chi_q}
\end{equation}

Since $\chi_q$ and $G_H$ have the same renormalization coefficient, the ratio $G_H/(2\chi_q T)$ does not require renormalization. On the lattice, however, due to charge conservation, $G_{00}(\tau, \vec{0})$ is essentially constant in $\tau$, except for the short-distance part of $G_{00}(\tau, \vec{0})$ related to cut-off effects. As a result, we calculate $G_{H}/(2\, G_{00}(\tau=1/2T, \vec{0}))$ for an estimate of $G_{H}/({2 \chi_q T})$. 

To perform the continuum extrapolation, we need lattice data for various lattice spacings at a fixed value of $\tau T$ and which may not be available for coarser lattices. To address this, we used cubic spline interpolation of the courser lattices to estimate the correlator at the same value of $\tau T$ for different lattice spacings. We determined the error on the interpolated data points through a bootstrap procedure. This involved performing the interpolation on each bootstrap sample. Once we estimated $G_H/(2\chi_q T)$ with the error at the same $\tau T$ for different lattice spacing, we performed the continuum extrapolation using the following ansatz:
\begin{equation}
\label{GH lattice}
\frac{G_{H}}{2 \chi_q T}(a)= \frac{G_{H}}{2 \chi_q T}(a=0)+\frac{b}{N_\tau^2} \,.
\end{equation}
We illustrated the fitting of the above ansatz for a few values of $\tau T$ in left panel of \Cref{cont_corr}.
Since this is a linear fit, the error bar on the continuum-extrapolated data points has been estimated using Gaussian error propagation. The continuum extrapolated correlator along with the correlator at finite lattice spacing is shown in the right panel of the same figure. We observe that the cut-off effect on this correlator at the momenta $k/T=1.57$ and $3.14$ is barely visible; however, with increasing momentum, cut-off effects start showing up when $\tau T \lesssim 0.25$. In contrast, the vector channel shows a significant cut-off effect due to the presence of a large UV component \cite{Ding:2016hua}.  This also indicates that T-L correlators are dominated by the IR part of the spectral function.

%% file: comp_with_pt.tex
\begin{figure*}
    \centering
    \includegraphics[width=8cm]{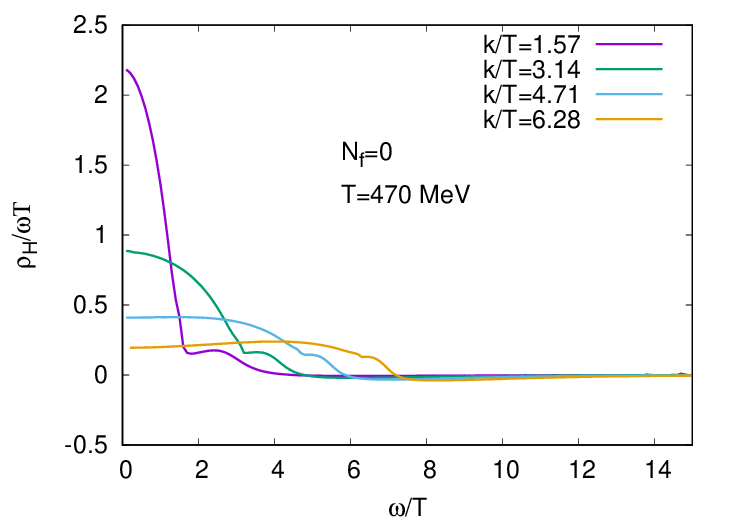}
    \includegraphics[width=8cm]{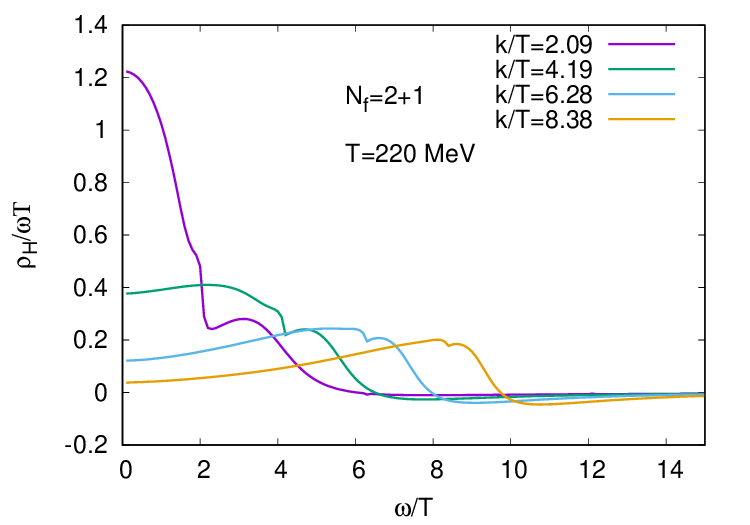}
	\caption{
	Perturbative spectral function in the T-L channel computed using the procedure from Ref.~\cite{Jackson:2019yao} at \NLPM, as a function of $\omega$ for several values of spatial momenta. ($\text{LPM}^\text{LO}$ means that near the light cone LPM resummation is performed at leading order.) Here the renormalization scale was fixed at $\mu = \mu_{\rm opt}$ as described in the main text. The left panel shows the spectral function with no dynamical quarks at $T=\SI{470}{\mega\eV}$, and the right panel shows the spectral function in $N_f=2+1$ flavor QCD at $T=\SI{220}{\mega\eV}\,$.}
    \label{per_sp}
\end{figure*}

\begin{figure*}
    \centering
    \includegraphics[width=8cm]{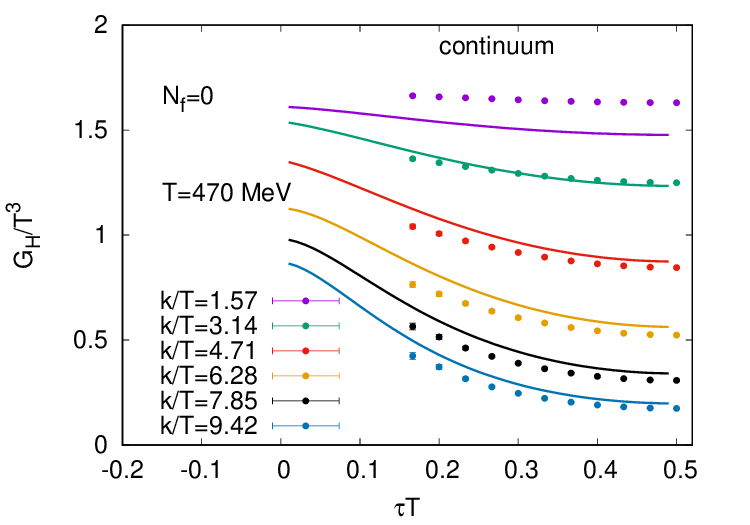}
    \includegraphics[width=8cm]{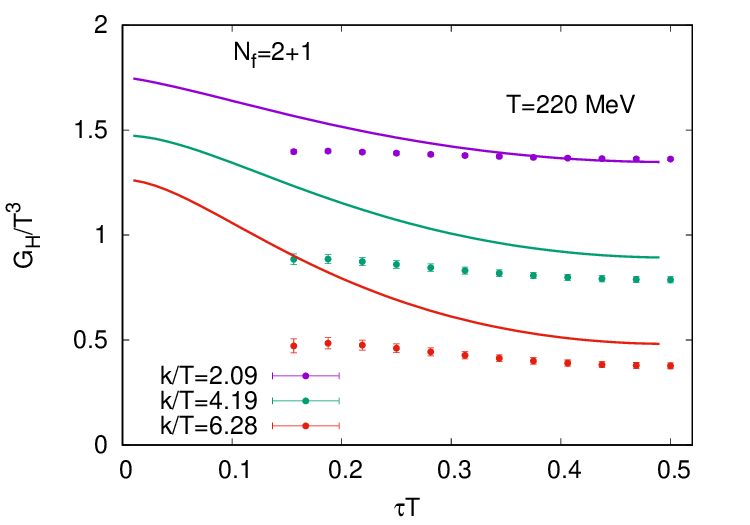}
	\caption{
	Euclidean correlator in T-L channel, calculated from the \NLPM perturbative spectral function and the corresponding lattice measurements for several momenta. 
	(As mentioned in the main text, the actual lattice data are multiplied by 
	$\chi_q=0.897\,T^2$ for the case of $N_f=0$ and $\chi_q=0.842\,T^2$ for $N_f=2+1$ in order to plot $G_H/T^3$.)
	Solid lines depict the perturbative results computed from \Cref{eq:KL}, while data points represent lattice correlators for quenched on the left and full QCD on the right.
	}
    \label{corr_per_latt}
\end{figure*}

Perturbation theory is a useful tool for calculating spectral functions, which becomes rigorous at high temperature thanks to asymptotic freedom, and significant progress has been made in using it for this purpose \cite{Laine:2013vpa,Jackson:2019mop,Jackson:2019yao,Arnold:2001ba,Aurenche:2002wq}. However, at the phenomenologically interesting temperature range, QCD is strongly coupled and weak-coupling calculations may not be sufficient. Therefore, non-perturbative calculations are necessary to access the photon rate. In this section, we compare the lattice correlator with the one obtained from the perturbative spectral function to judge the importance of non-perturbative effects.

At leading-order (LO), the spectral function is determined from the process $q\bar q\to\gamma^*$. At next-to-leading-order (NLO), the spectral function receives contributions from well-known processes such as Compton scattering ($qg\to\gamma^*q$), pair annihilation ($q\bar q\to\gamma^*g$), various interference terms between them,
as well as virtual loop corrections to the LO process~\cite{Baier:1988xv,Gabellini:1989yk,Altherr:1989jc}. 
On the light cone, the LO spectral function is zero, but at NLO, it exhibits a well-known logarithmic singularity and a finite discontinuous part. 
Far away from the light cone ($M^2=|\omega^2-k^2|\gsim(\pi T)^2$), perturbative calculations are well-behaved~\cite{Laine:2013vpa,Jackson:2019mop}. 
However, the singularity of the NLO spectral function at the light cone indicates the breakdown of the naive perturbative calculation, necessitating resummation to all orders of a certain class of diagrams~\cite{Braaten:1990wp,Moore:2006qn}. 
Therefore, a separate treatment is required to calculate the spectral function close to the light cone.

When $M^2\lsim g^2 T^2$, the hard thermal loop (HTL) resummation introduces an asymptotic (thermal) mass on the internal quark propagator in $2\to2$ NLO processes, which regulates the logarithmic singularity at the light cone~\cite{Kapusta:1991qp,Baier:1991em}. 
Additionally, multiple scatterings due to soft gluons can occur with the quark that emits the photon.\footnote{These interactions can happen within the formation time of the photon and end up  contributing significantly to the rate. It is also mandatory to resum such scatterings beyond leading-logarithmic accuracy.}
To include this contribution, a technique called Landau-Pomeranchuk-Migdal (LPM) resummation is used in this regime~\cite{Arnold:2001ba, Aurenche:2002wq}. 
This resummation removes the singularity and the discontinuity of the spectral function at the light cone and therefore makes the spectral function smooth across the light cone.

We computed the perturbative spectral function and its corresponding correlator 
for the temperatures and momenta used in our lattice study, following the calculation described in Ref.~\cite{Jackson:2019yao}. In that paper, leading order LPM resummation near $M^2\sim g^2 T^2$ was performed by solving a two-dimensional Schrödinger equation with a LO light cone potential \cite{Laine:2013lia}. The spectral function in the intermediate region is obtained by interpolating between the $M^2\sim T^2$ and $M^2 \sim g^2 T^2$ region. The renormalization scale chosen for the calculation is $\mu_{\rm opt}=\sqrt{M^2+(\zeta \pi T)^2}$ 
with $\zeta^{(N_f=0)}=1$ and $\zeta^{(N_f=3)}=2\,$. 
This choice is motivated by the fact that, far from the light cone, the relevant scale is $K^2 = \omega^2 - k^2$ (the photon's virtuality), whereas near the light cone, the physics is that of  dimensionally reduced EQCD, and thus the relevant scale is temperature~\cite{Caron-Huot:2008zna}. The running coupling then has been calculated by using $\Lambda^{{\overline{\rm MS}}}=0.8\, T_c$ for quenched~\cite{Francis:2015lha} and  $\Lambda^{{\overline{\rm MS}}}=339$~MeV~$\simeq1.88\, T_{pc}$ for full QCD~\cite{FLAG} (these choices have uncertainties $\sim 10 ... 25$\%).

The resulting spectral function is plotted for various momenta in \Cref{per_sp} for both quenched and full QCD. As expected, the spectral function is UV suppressed, and it has been shown~\cite{Jackson:2019yao} that it indeed behaves as $1/\omega^4$ for $\omega\gg k,\pi\,T$ region, which is in agreement with the OPE expectations and satisfies the sum rule~\cite{Caron-Huot:2009ypo}.

As discussed in the previous section, the ratio $G_{H}/(2\chi_q T)$ on the lattice is free of renormalization. In order to compare it with the perturbative correlator, we used $\chi_q=(0.897\pm0.012) \,T^2$ for the case of $N_f=0$ and $\chi_q=0.842^{+0.05}_{-0.03} \,T^2$ for $N_f=2+1$. The value for $N_f=0$ was estimated using the Lüscher non-perturbative method ~\cite{Ding:2016hua}, while for the case of $N_f=2+1$, we used the perturbative EQCD estimate of order $g^6 \log(g)$ \cite{Vuorinen:2002ue}. 
The errorbar on the non-perturbative estimate is statistical in nature ~\cite{Ding:2016hua}, whereas for the perturbative estimate, the error bar is obtained by varying the renormalization scale between $\mu=0.5\,\mu_{\rm min}$ to $\mu=2.0\,\mu_{\rm min}$, where $\mu_{\rm min}=9.082\,T$ is the minimum sensitivity scale. For $N_f=3$, $\chi_q$ for the physical pion mass was calculated in Ref.~\cite{Borsanyi:2011sw}, which provides an estimate $\chi_q=(0.772 \pm 0.015)\,T^2$ at the temperature $1.2T_c$. In Ref.~\cite{Mogliacci:2013mca}, a {\em hard thermal loop} (HTL) resummed perturbative estimate gives a susceptibility $\chi_q=0.888^{+0.023}_{-0.063}\,T^2$. However, using these values for the $\chi_q$, the qualitative conclusion remains unchanged. We should also like to mention we need $\chi_q$ only to compare the perturbative correlator. The $D_{\text{eff}}$ can be directly calculated from the correlator $G_H / (2\chi_q T)$.
The comparison between the correlator obtained from the perturbative spectral function and the lattice calculation is illustrated in \Cref{corr_per_latt}.
 
 In the left panel, we plotted the quenched case correlator up to a large value of momentum, since the quenched results are continuum extrapolated. However, for full QCD, the results are only provided for the first three lattice momenta, as the lattice finite spacing effects are larger for higher momenta. In the quenched case, we observe that our lattice correlator data significantly exceeds the perturbative correlator at the lowest available momentum. However, at higher momentum, the situation improves, although a small non-perturbative difference remains. On the other hand, in full QCD, non-perturbative effects are already prominent for all  momentum values we considered. The temperatures for the $N_f=0$ and $N_f=2+1$ cases are $1.5\,T_c$ and $\sim 1.2\,T_{pc}$ respectively. The $N_f=0$ case temperature in this ``scaled temperature'' unit
 is $25\%$ higher than the $N_f=2+1$ case. Therefore, \Cref{corr_per_latt} also illustrates that by going $25\%$ higher in temperature, one sees a significant suppression in the non-perturbative effects.

%% file: spectral_reconstruction.tex
\begin{figure*}
    \centering
    \includegraphics[width=8cm]{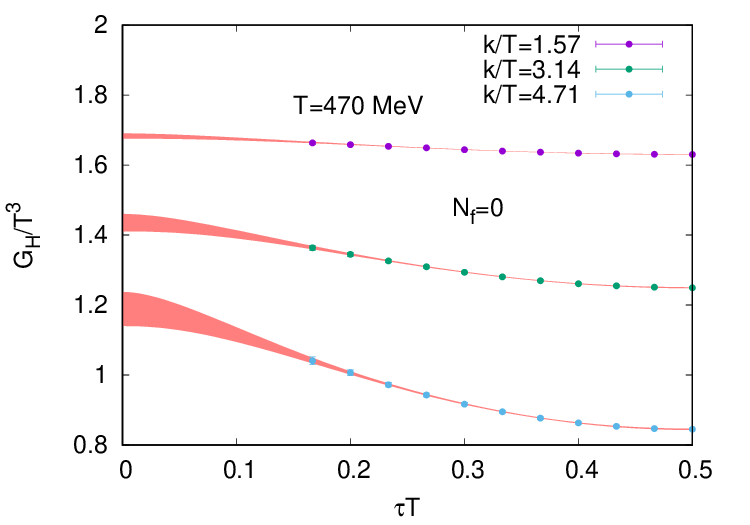}
    \includegraphics[width=8cm]{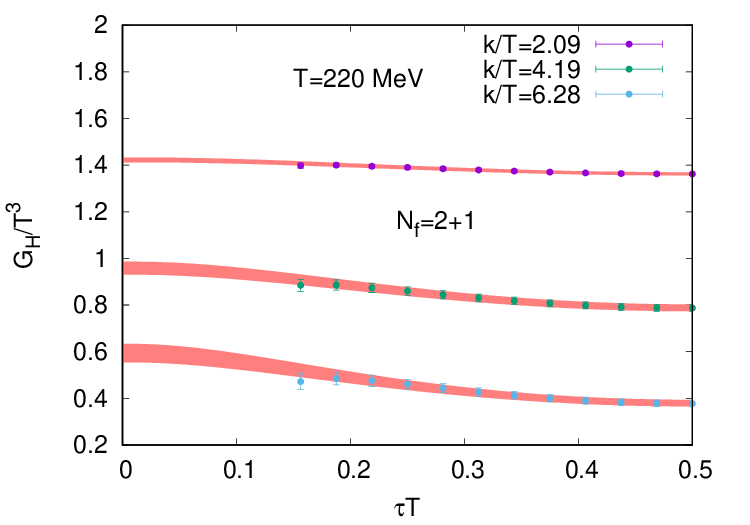}
	\caption{The Euclidean correlator in the T-L channel, illustrating the result of fitting the polynomial ansatz in \Cref{poly-ansatz} for $\omega_0=\sqrt{k^2+(\pi T)^2}$ (red band) to the measured lattice data. The quenched case is shown on the left and full QCD is shown on the right, both as a function of $\tau$ for just a few indicative momenta.}
    \label{fit_polynomial}
\end{figure*}

In this section, we will discuss some approaches to spectral reconstruction from the lattice Euclidean correlator.
Despite the ill-conditioned nature of the inverse problem, one can make use of additional, physically motivated, information to improve the reliability of extracted spectral functions---a strategy which has been successful in previous works~\cite{Gupta:2003zh, Ghiglieri:2016tvj,Brandt:2017vgl,Ce:2020tmx}.
As discussed in \Cref{sec:t-l_spectral}, the known physical constraints on the spectral function contain the sum rule \Cref{eq:sumrule} and the asymptotic behavior $1/\omega^4$ at $\omega\gg k,\pi T$. In the following, we discuss several methods and models for the spectral reconstruction that fold in these physical constraints \cite{Bala:2022qib}. 

\subsection{Models of the spectral function}

Firstly, we consider a model of the T-L spectral function 
in which the $\omega$-dependence arises from connecting 
two regions (delineated by 
$\omega_0\gsim k,\pi\,T$). 
The `ultraviolet' region $\omega \ge \omega_0$ is modeled by $\rho_>$ 
which is constructed from inverse even powers of $\omega$ (starting with $1/\omega^4$)
in order to satisfy the OPE result. 
The `infrared' region $\omega \le \omega_0$ is modeled by $\rho_<$, 
for which we adopt the same polynomial put forward in Ref.~\cite{Ghiglieri:2016tvj}.
The two regions are matched continuously, 
\begin{equation}
     \rho_{H}^{\rm poly}(\omega) \; = \; 
     \rho_<(\omega) \, \Theta(\omega_0 - \omega) +
     \rho_>(\omega) \, \Theta(\omega - \omega_0) \, , \label{poly-ansatz}
\end{equation}
where $\Theta$ is the Heaviside step function.
Because the spectral function is expected to be smooth and differentiable across the light cone~\cite{Caron-Huot:2008zna}, 
we introduce the parameters $\beta$ and $\gamma$ such that 
\begin{align}
    \rho_<(\omega_0) &= \; \rho_>(\omega_0) \; = \; \beta \, , \\
    \rho^\prime_<(\omega_0) &= \; \rho^\prime_>(\omega_0) \; = \; \gamma \, .
\end{align}

These considerations lead to the following model of the spectral function
\begin{widetext}
\begin{align}
    \rho_< &\equiv
    \frac{\beta \omega^3}{2\omega_0^3}\left(5-3\frac{\omega^2}{\omega_0^2}\right)
    -\frac{\gamma \omega^3}{2\omega_0^2}\left(1-\frac{\omega^2}{\omega_0^2}\right)
    +\frac{\delta \,\omega}{\omega_0} \left(1-\frac{\omega^2}{\omega_0^2}\right)^2\,, \\[1ex]
    \rho_> &\equiv
    -\frac{\beta\, \omega_0^4}{7\,\omega^4}\left (54 \frac{\omega_0^4}{\omega^4}\,-\,94 \frac{\omega_0^2}{\omega^2}\,+\,33\right)
    +\frac{\gamma \, \omega_0^5}{140\,\omega^4}\left (-81 \frac{\omega_0^4}{\omega^4}\,+\,92 \frac{\omega_0^2}{\omega^2}\,-\,11\right)
    -\frac{16\,\delta\,\omega_0^4}{35\,\omega^4}\left(1-\frac{\omega_0^2}{\omega^2}\right)^2
     \,,
\end{align}
\label{polyeq}
\end{widetext}
\begin{figure*}
    \centering
    \includegraphics[width=8cm]{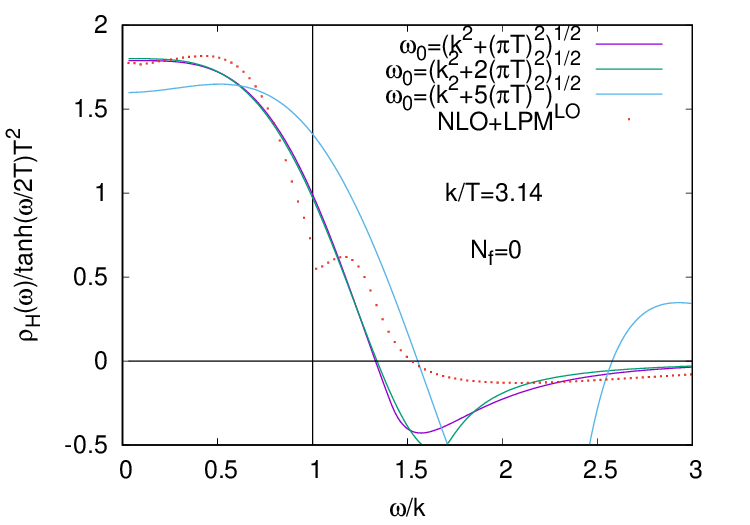}
    \includegraphics[width=8cm]{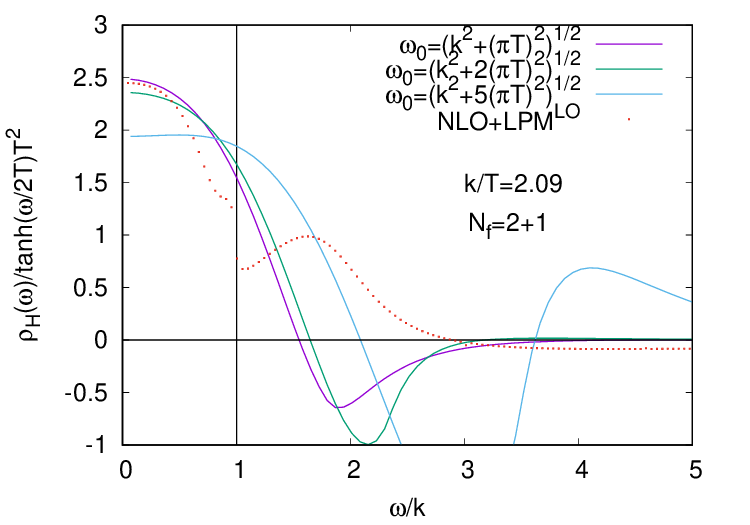}
    \includegraphics[width=8cm]{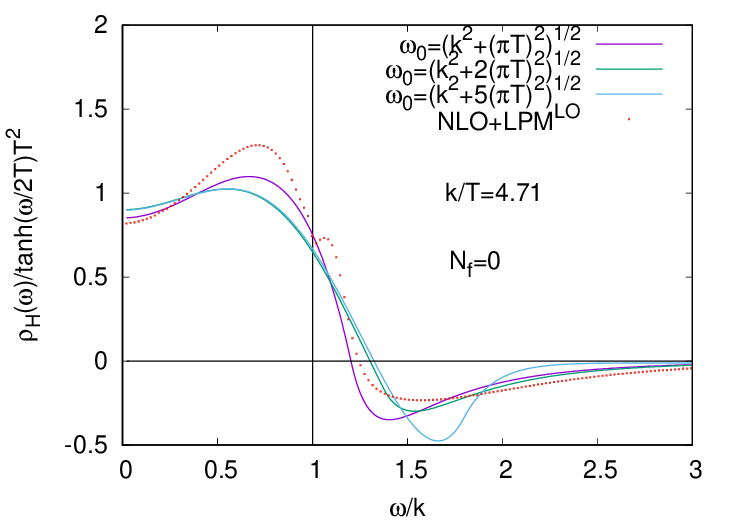}
    \includegraphics[width=8cm]{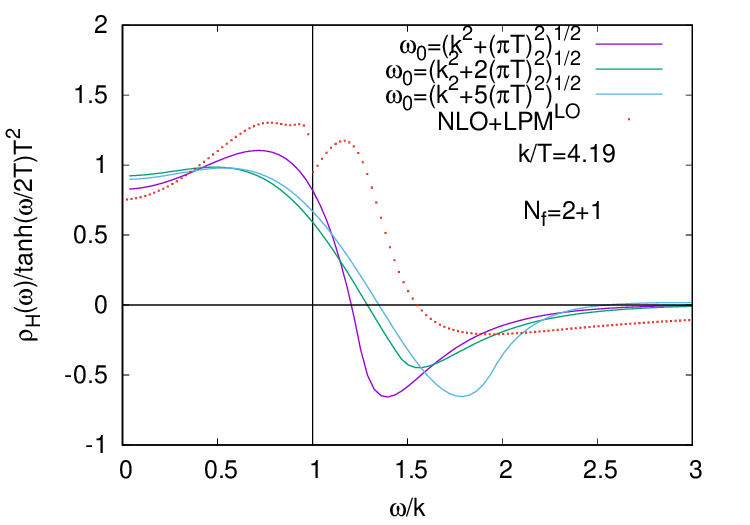}
    \includegraphics[width=8cm]{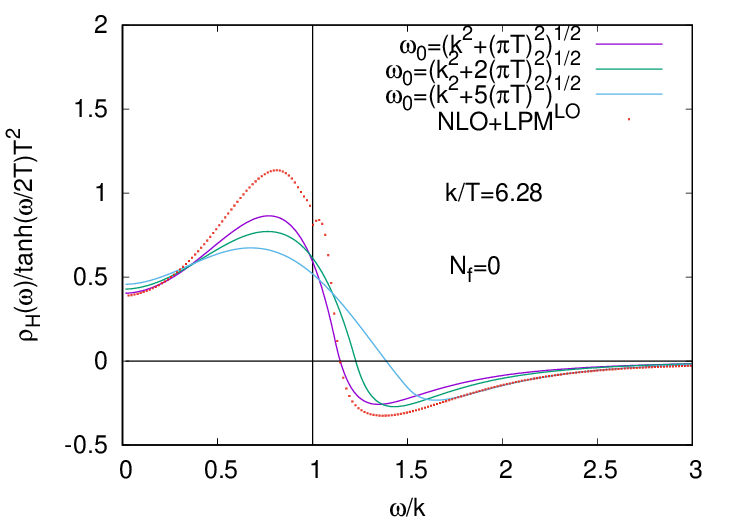}
    \includegraphics[width=8cm]{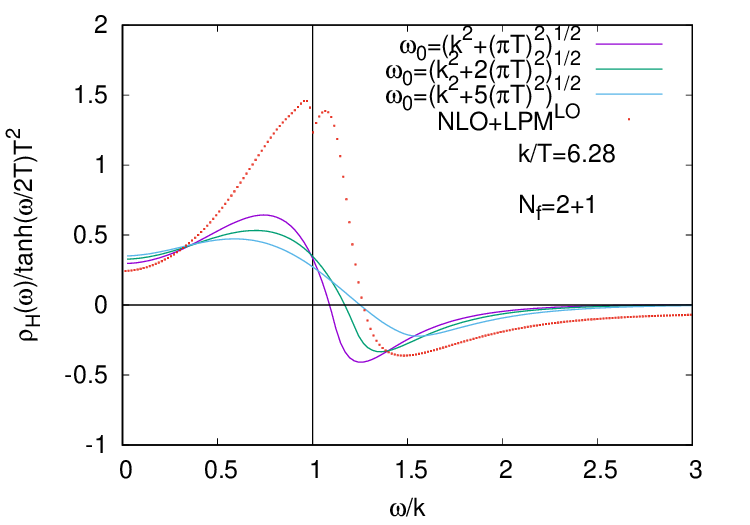}
	\caption{
	The spectral function $\rho_H$ for various momenta obtained from the polynomial fitting (solid lines), 
	and the dependence on the matching point $\omega_0$ is also illustrated. 
	The results for quenched QCD are shown on the left, while the results for full QCD are shown on the right.
	The perturbative version of the same spectral function is also shown for $\mu=\mu_{\rm opt}$ (red points).
	}
    \label{Quenched_polynomial_sp}
\end{figure*}
where the coefficients have been chosen to ensure the matching. 
The quantity $\delta$ is another parameter which controls the slope of the spectral function at $\omega=0$. 
The parameters $\beta$, $\gamma$, and $\delta$ are chosen such that the sum rule is satisfied.
The fitting parameters are constrained to values that satisfy $\delta > 0 $, $\rho_{H}(k,k) > 0$ and $\frac{\partial G_H}{\partial \tau}\le 0$. The resulting fits of the lattice data with this model of the spectral function is shown in \Cref{fit_polynomial} for quenched and full QCD, while the spectral function obtained from this fit is illustrated in \Cref{Quenched_polynomial_sp}. 

We estimate the error of the reconstruction by comparing different values of $\omega_0=\sqrt{k^2+\nu\,(\pi T)^2}$ for $\nu=\{1,2,5\}$. In \Cref{Quenched_polynomial_sp} we show the variation of the spectral function for 
a these values of $\omega_0$.
At small momentum, we see that for both quenched and full QCD, the variation with respect to $\omega_0$ becomes significant only at large $\omega_0$. The polynomial spectral function also predicts a large negative part at large $\omega$ compared to the perturbative spectral function. At higher momentum, however, we see that the variation of the spectral function with respect to $\omega_0$ becomes small for quenched QCD. The large $\omega$ part of the spectral function is also approximately consistent with the perturbative spectral function. At the same time, for full QCD, we see that the variation with respect to $\omega_0$ is comparatively larger than the quenched case, and the polynomial model spectral function shows a large deviation from the corresponding perturbative spectral function. We pursue a mock analysis of the perturbative correlator in \Cref{app:mock_poly}, showing that this range of $\omega_0$ approximately captures the perturbative spectral function.

The photon production rate, which is proportional to $D_{\text{eff}}$, can then be obtained from the spectral function. The systematic error on $D_{\rm eff}$ was obtained by combining the values of $\omega_0$ prescribed
by $\nu = \{ 1,2,5\}\,$. 
We provide the resulting numerical values for $D_{\text{eff}}$ in \Cref{Tab:Deff quenched} for quenched and in \Cref{Tab:Deff 3 flavour} for full QCD. 
The results shown are from continuum extrapolated data set for quenched QCD. In \Cref{app:finite_latt_spacing}, we discuss the dependence of the $D_{\text{eff}}$ on the lattice spacing.

\begin{figure*}
    \centering
    \includegraphics[width=8cm]{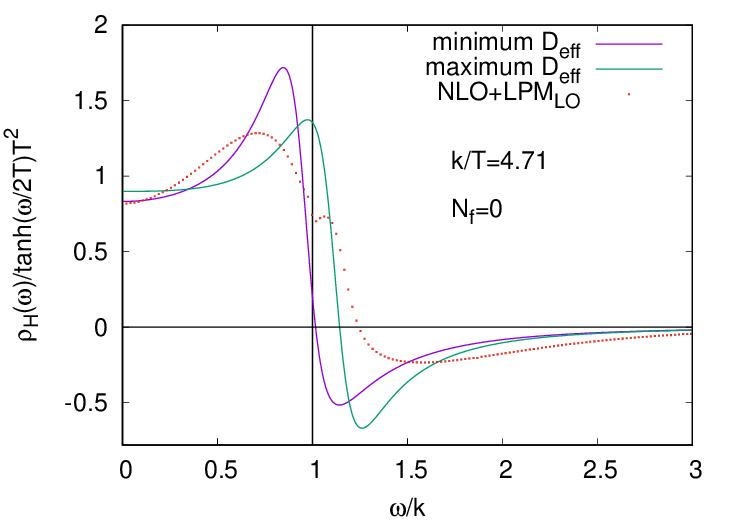}
    \includegraphics[width=8cm]{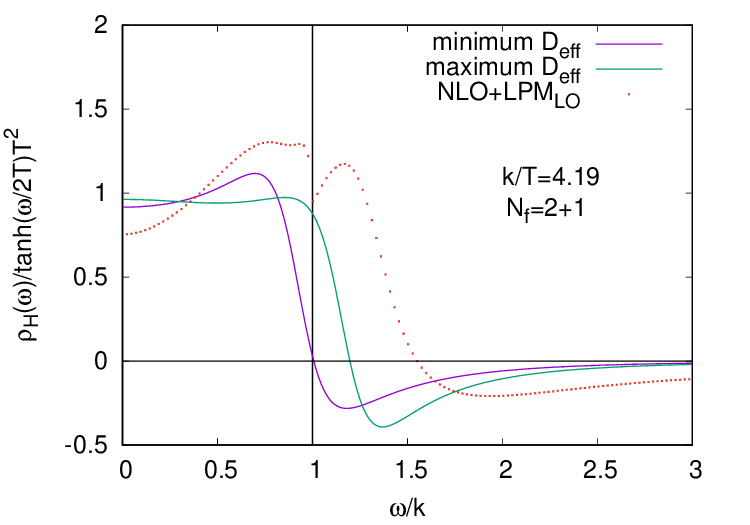}
	\caption{
	Spectral function $\rho_H$ obtained from the Pad\'e-like ansatz in \Cref{Pade}, after fitting to the corresponding lattice Euclidean correlator.
	On the left, the spectral functions are shown for quenched QCD at momentum $k/T=4.71$, and on the right, for full QCD at momentum $k/T=4.19$,
	with the minimum $D_{\text{eff}}T$ and maximum $D_{\text{eff}}T$ samples of the bootstrap distribution.}
    \label{pade_sp}
\end{figure*}

The second model of the T-L spectral function is a Pad\'e-like ansatz that has 
already been applied to the reconstruction problem in Ref.~\cite{Ce:2020tmx} 
and is given by
\begin{align}\label{Pade}
    \rho^\text{Pad\'{e}}_{H}(\omega)=\frac{A \tanh(\frac{\omega}{2 T}) \big(1+B \omega^2\big)}{(\omega^2+a^2)\big[(\omega+\omega_0)^2+b^2\big]\big[(\omega-\omega_0)^2+b^2\big]}
    \,.
\end{align}
At small $\omega$ this spectral function reproduces the hydrodynamic prediction 
$\rho_{H}(\omega)\sim \frac{A \omega}{\omega^2+a^2}$ for the IR limit~\cite{Meyer:2011gj}. 
The remaining part of the spectral function is inspired by qualities from AdS/CFT (see, e.g. Ref.~\cite{Kovtun:2005ev}) and at large $\omega$ is consistent with the OPE. 
In order to satisfy the sum rule, $B$ becomes a function of $\omega_0,\,a$ and $b$ 
(here the parameter $\omega_0$ is not the same as the one from \Cref{poly-ansatz}). 
The fit then was performed with respect to $A,\,a,\,b,\,\omega_0$ by minimizing the uncorrelated $\chi^2$. 
We perform this fit in each of the bootstrap samples. 
The resulting distribution of $D_{\text{eff}}T$ over the bootstrap samples is broad, indicating a near-degenerate minimum $\chi^2$, as also identified in Ref.~\cite{Brandt:2017vgl}. 
Therefore, we quote the variation of the minimum and maximum $D_{\text{eff}}T$ of the distribution within the error bars (rather than the standard deviation from the mean values). 
The estimated $D_{\text{eff}}$ can be found in \Cref{Tab:Deff quenched} for quenched and \Cref{Tab:Deff 3 flavour} for full QCD. 
Sample spectral functions for the two extreme cases are plotted in \Cref{pade_sp}, along with the perturbative spectral functions. We see that the qualitative behavior of these spectral functions are similar to the spectral function obtained from the polynomial ansatz. However at large $\omega$ the Pad\'e-like ansatz shows a large negative contribution compared to the perturbative spectral function.

%

\subsection{Backus-Gilbert Estimate}
%
\begin{figure*}
    \centering
    \includegraphics[width=8cm]{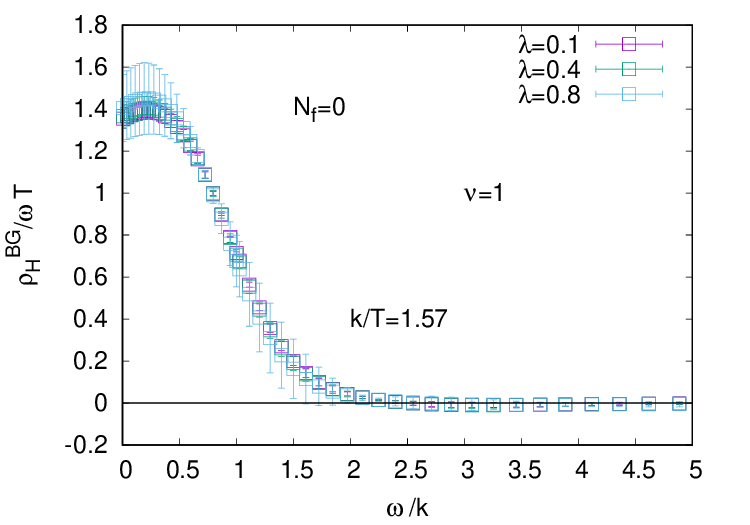}
    \includegraphics[width=8cm]{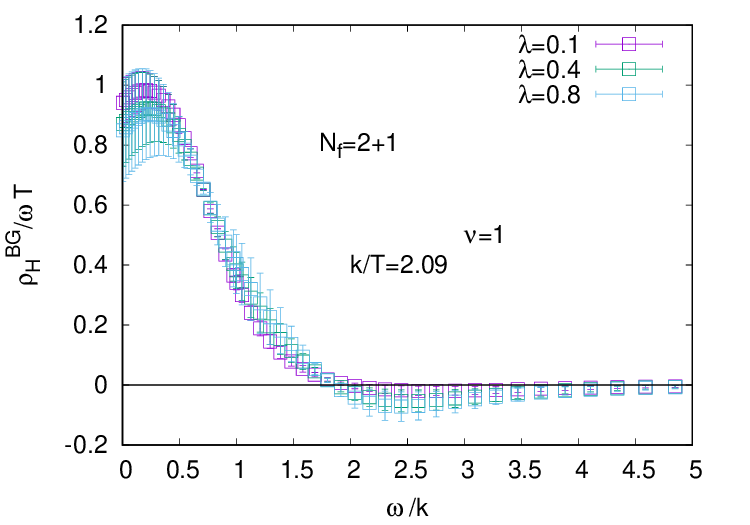}
	\caption{
	Backus-Gilbert estimated spectral function, with different colored points representing different choices of $\lambda$, the regularizing parameter.
	The same $\nu=1$ is used for both quenched (left) and full QCD (right), corresponding to 
	$\omega_0 = \sqrt{k^2 + (\pi T)^2}$ in the prior function \Cref{priorbg}. 
	}
    \label{spfn_lambda}
\end{figure*}

\begin{figure*}
    \centering
    \includegraphics[width=8cm]{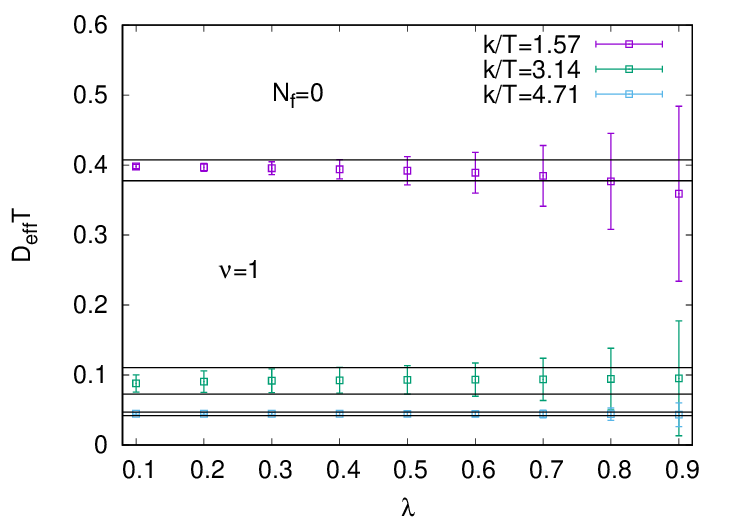}
    \includegraphics[width=8cm]{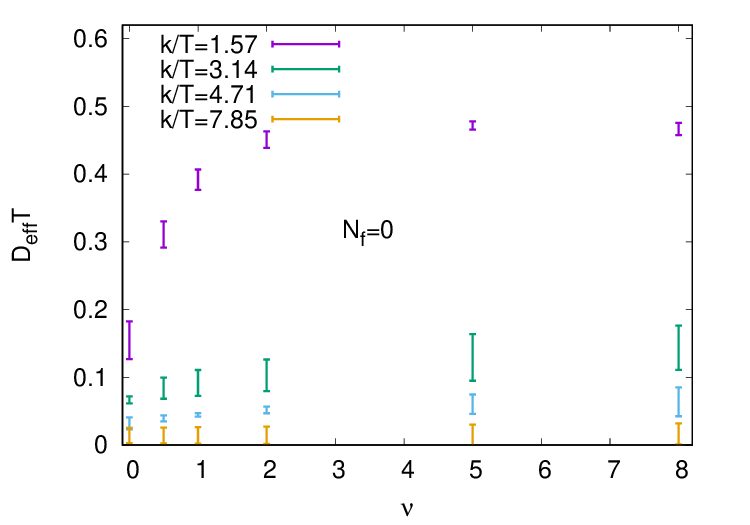}
    \includegraphics[width=8cm]{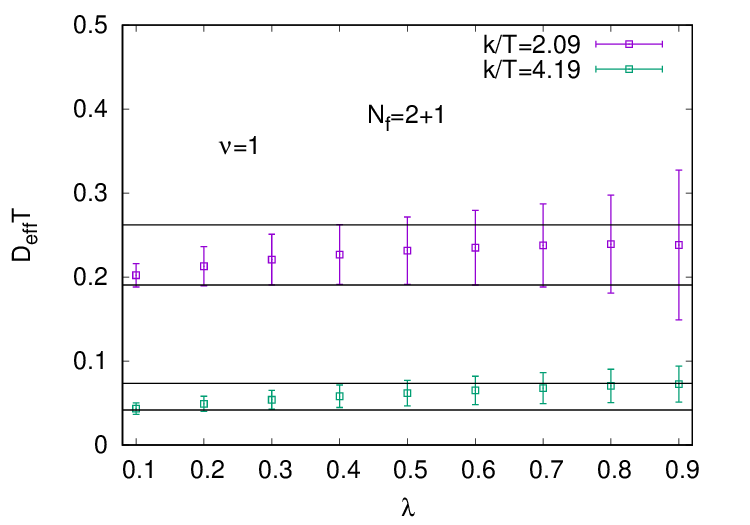}
    \includegraphics[width=8cm]{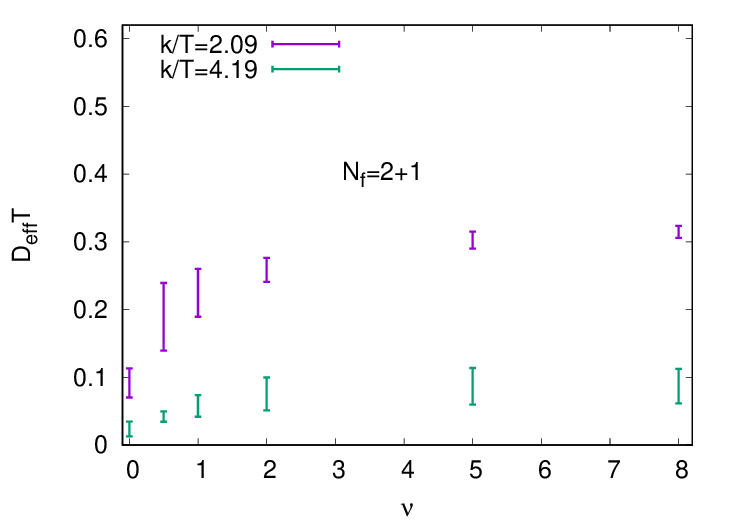}
	\caption{
	Dependence of the diffusion coefficient $D_{\text{eff}}T$, from the BG-estimated spectral function, on the regularization parameter $\lambda$ for $\nu = 1$ is shown on the left and
	of the bootstrap average (indicated by black horizontal lines). The same dependence on $\nu$ appears on the right. 
	The top two panels are for $N_f =0$ and the lower two panels are for $N_f = 2+1\,$, in each case several momenta are shown.
	}
    \label{Deff_lambda}
\end{figure*}

\begin{figure*}
    \centering
    \includegraphics[width=8cm]{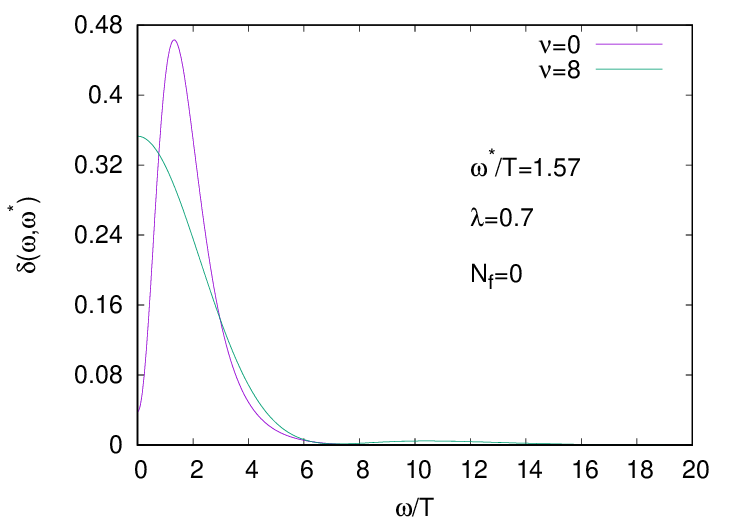}
    \includegraphics[width=8cm]{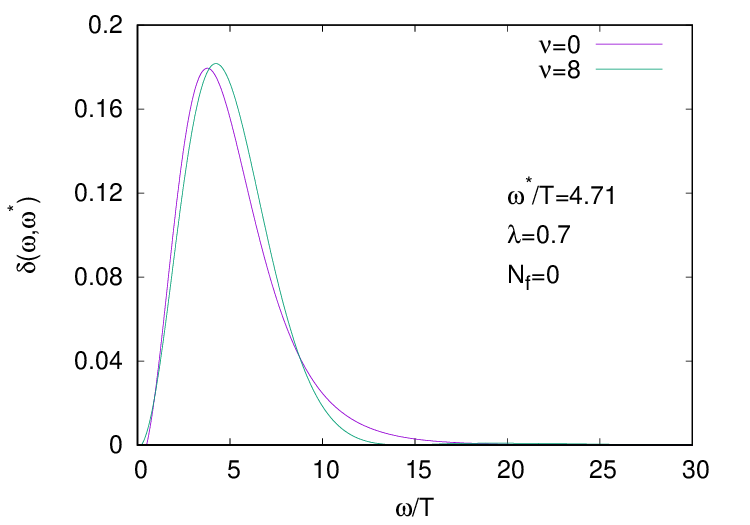}
	\caption{
	Dependence of the resolution function $\delta(\omega, \omega^* = |\vec k|)$ 
	on $\omega\,$, for parameter values of $\nu = \{ 0, 8 \}$ and at $\lambda = 0.7\,$.
	Both panels are for quenched QCD, the left for $k/T=1.57$ and the right for $k/T=4.71\,$.
	}
    \label{res_func}
\end{figure*}

The Backus-Gilbert (BG)~\cite{Backus:1968svk} method is a technique commonly used for spectral reconstruction in the context of QCD. It involves computing a smeared estimator for the spectral function, denoted as $\rho^{\rm BG}(\omega, \vec k)$. Consequently, it represents the actual spectral function $\rho(\omega, \vec k)$ convoluted with a so-called resolution function $\delta (\omega,\omega^{*})$, namely
\begin{equation}
    \rho^{\rm BG}(\omega, \vec{k}) = \int {\rm d}\omega^{*}  \, \delta(\omega, \omega^{*})\,  \rho(\omega^{*}, \vec{k})
    \,.
\end{equation}
When the resolution function $\delta (\omega,\omega^{*})$ is a Dirac delta function $\delta (\omega-\omega^{*})$, the Backus-Gilbert estimate provides an exact reconstruction of the spectral function. However, due to the problem at hand being ill-conditioned, this is not possible in practice. 
Therefore, the objective of this method is to minimize the width of the resolution function in order to achieve better accuracy.

In order to identify the resolution function in this method, the correlator in \Cref{eq:KL} is rewritten as
\begin{equation}\label{eq:BG}
    G_{H}(\tau, \vec k) \; = \;
    \int_{0}^{\infty} {\rm d}\omega \,
    \frac{\rho(\omega,\vec k)}{ f(\omega,\vec k)} \, 
    \tilde{K}(\omega,\tau) 
    \,,
\end{equation}
where $\tilde{K}(\omega,\tau)\equiv\frac{f(\omega,\vec k) \cosh[\omega(1/2T-\tau)]}{\sinh(\omega \beta/2)\pi}$.
Here, some arbitrary function $f(\omega,\vec k)$ is introduced by hand. It can be chosen to incorporate the known physics information about the spectral function asymptotics and should also remove the singularity at $\omega = 0$ originating from the kernel.

The BG-estimated spectral function is expressed as a linear combination of the measured lattice correlator points
\begin{widetext}
\begin{equation}
    \frac{\rho_{H}^{\rm BG}(\omega,\vec k)}{f(\omega,\vec k)}
    \; = \; 
    \sum_{i} q_{i}(\omega,\vec k) \,G_{H}(\tau_{i},\vec k)\\
    \; = \;
    \int_{0}^{\infty} {\rm d}\omega^{*} \,
    \delta(\omega^{*},\omega) \,
    \frac{\rho_{H}(\omega^{*},\vec k)}{f(\omega^{*},\vec k)}
    \,,
\end{equation}
\end{widetext}
where we identify $\delta(\omega,\omega^{*})=\sum_{i} q_{i}(\omega,\vec k) \tilde{K}(\omega^*,\tau_i)$. 

For any $\omega$, the resolution function is therefore fully characterized by the coefficients $q_i$. One way of constructing these coefficients, is by minimizing $A(\omega)$ 
(w.r.t. $q_i$), 
a measure of the width of the resolution function
\begin{equation}
    A(\omega) \; \equiv \;
    \int {\rm d} \omega^*\,(\omega- \omega^*)^2 \delta(\omega, \omega^*)^2
    \; = \;
    {\boldsymbol{q}^\intercal(\omega)\boldsymbol{W}(\omega)\boldsymbol{q}(\omega)}
    \,,
\end{equation}
where  
 $[\boldsymbol{W}]_{ij}(\omega)\equiv \int_0^{\infty} {\rm d} \omega^* \, (\omega-\omega^*)^2 \tilde{K}( \omega^*,\tau_i) \tilde{K}( \omega^*, \tau_j)$.

In the BG method, the minimization is performed under the constraint that the integral of the resolution function $\delta(\omega, \omega^{*})$ over $\omega$ is equal to 1.
With this constraint, the solution for the coefficients 
$[\boldsymbol{q}]_i$ 
is given by
\begin{equation}
    \boldsymbol{q}(\omega) \;=\; 
    \frac{\boldsymbol{W}^{-1}(\omega)\, \boldsymbol{r}}{ \boldsymbol{r}^\intercal \, \boldsymbol{W}^{-1}(\omega)\, \boldsymbol{r}}
    \,,
\end{equation}
where 
$[\boldsymbol{r}]_i\equiv\int {\rm d}\omega\,\tilde{K}(\omega,\tau_i)\,$.
In reality, the matrix $\boldsymbol{W}$ is close to singular---due to its large condition number---and therefore  requires regularization.

One of the popular regularization schemes is to take
\begin{equation}
    \boldsymbol{W}^{\lambda} \; \equiv\; \lambda \, \boldsymbol{W} + (1-\lambda) \, \boldsymbol{S}
    \,,
\end{equation}
where $\boldsymbol{S}$ is the covariance matrix of the lattice correlator. Other regularization schemes also exist in literature, e.g. the Tikhonov regularization \cite{Astrakhantsev:2018oue} where $\boldsymbol{S}$ is replaced by the identity matrix. In this paper, we have used a diagonal covariance matrix for $\boldsymbol{S}$. 

The chosen regularization introduces an additional dependence of the spectral function on the parameter $\lambda$. The spectral function also depends on the prior function. We choose the form of the prior function as
\begin{equation}\label{priorbg}
   f(\omega ,\vec k)=\left(\frac{\omega_0}{\omega}\right)^4 \tanh\left(\frac{\omega}{\omega_0}\right)^5
   \,,
\end{equation}
where $\omega_0=\sqrt{k^2+\nu (\pi T)^2}$ as before.
In this case, we will vary $\nu = \{ 0, \frac12, 1, 2, 5, 8 \}\,$. 
The functional form is  based on the information that we already have about the spectral function, i.e it behaves as $1/\omega^4$ for large $\omega \gg k $ values consistent with the OPE expectation, and linearly with $\omega$ for small $\omega$ values.

The resulting BG-estimated spectral function, and thus the photon production rate, depend on the parameters $\nu$ and $\lambda$. The BG-estimated spectral functions for the momentum $k/T=1.57$ (for $N_f=0$) and $k/T=2.09$ (for $N_f=2+1$) are shown in \Cref{spfn_lambda}. In this figure, we showed the dependence of spectral function on $\lambda\in[0.1,0.8]$  for $\nu=1\,$. 
(The errors on the spectral functions are statistical.)
Here we also observed the general feature, that the UV part of the spectral function is  suppressed compared to the IR part. 
The photon production rate is proportional to the diffusion coefficient $D_{\text{eff}}$, 
defined from $\rho_H$ in \Cref{eq:Deff}, which is plotted on the left panel of \Cref{Deff_lambda} as a function of $\lambda$ for various momenta, but for a given value of $\nu$. 
We observe that for both $N_f=0$ and $N_f=2+1$ case the estimated $D_{\text{eff}}$ is stable with respect to the variation of the $\lambda$ parameter, although the error starts growing larger as $\lambda \to 1\,$. 
The values of  $D_{\text{eff}}$ for various $\lambda$ are combined using a bootstrap average over the plateau in region $\lambda= [ 0.1 , 0.7]$ in steps of $\Delta \lambda=0.1 $.
This $\lambda$-averaged version of $D_{\text{eff}}T$ is plotted as a function of $\nu$ on the right panel of \Cref{Deff_lambda}. 
For both $N_f=0$ and $N_f=2+1\,$, the coefficient $D_{\text{eff}}T$ exhibits small variations with respect to $\nu\,$, except for the smallest cases displayed, where deviations emerge for $\nu \lsim 2\,$.

The reason for this can be understood by looking at the resolution function, 
which is depicted in \Cref{res_func}. 
In this plot, we show the resolution function for the momentum $k/T=1.57$ and $k/T=4.71$ 
for the minimum and maximum $\nu$ values in the case of $N_f=0$ and $\lambda=0.7\,$. 
We find that for $k/T=4.71$, 
the resolution function is peaked around the light cone and exhibits variations of $\sim 9\%$ from $\nu=0$ to $\nu=8$. 
As a result, the BG-estimated spectral function in this case gets most of its contribution from the light cone region. 
Apart from the lowest momentum, this pattern is repeated for the other momenta studied here.
For $k/T=1.57$ and $\nu=0$, we find that the resolution function is peaked at the light cone. 
However, for $k/T = 1.57$ and $\nu=8$, the resolution function instead shows a peak at $\omega=0\,$. 
Therefore, the BG-estimated photon production rate from these values of $\nu$ gets a significant contribution from the $\omega=0$ region. Nevertheless, when quoting the value of the photon production rate, we included these variations with respect to $\nu$ as a source of systematic error.
In the case of full QCD, we find the results for the momentum $k/T=6.28$ are negative up to $\nu=2$. 
The tabulated values of $D_{\text{eff}}T$, determined from the BG method, 
can be found in \Cref{Tab:Deff quenched} for quenched and \Cref{Tab:Deff 3 flavour} for full QCD. 
This implies a lower bound for $D_\text{eff}$ equal to zero for the photon production rate because a negative rate would be unphysical. 
%

\subsection{Gaussian Process Regression}
%
\begin{figure*}
    \centering
    \includegraphics[width=8cm]{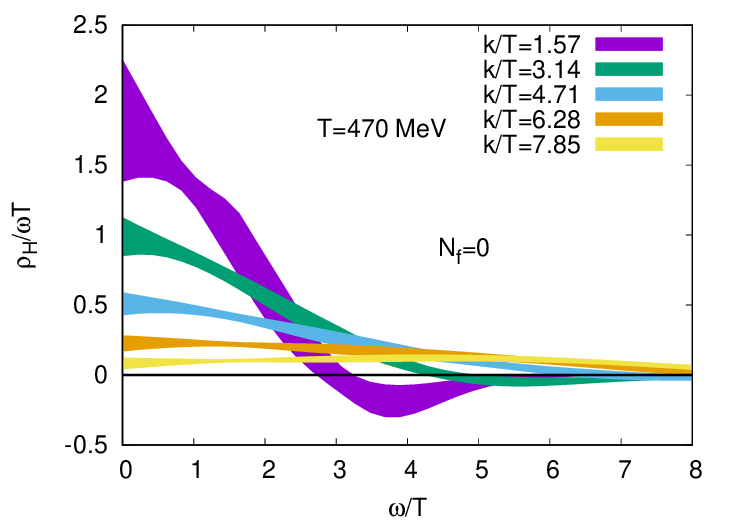}
    \includegraphics[width=8cm]{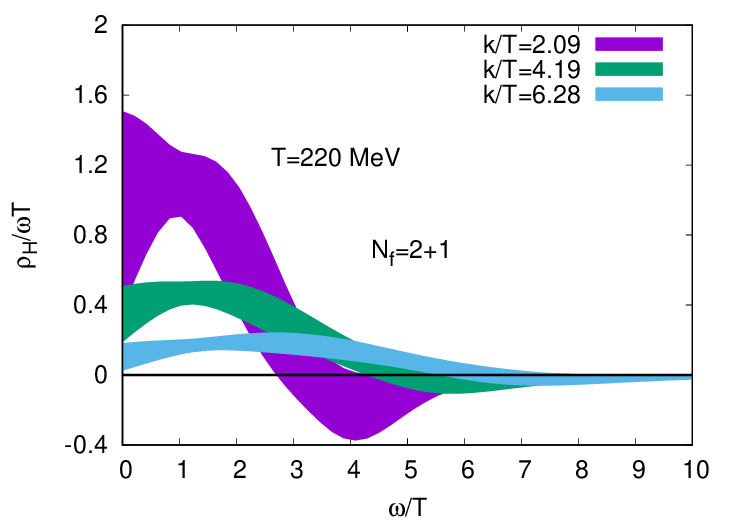}
	\caption{
	Spectral function from GPR: Quenched on the left, and full QCD on the right. The error band includes both the error propagation of the lattice data and the systematic error from the GP model.}
    \label{gp_spectral}
\end{figure*}

Gaussian Processes Regression (GPR), also known as `Kriging' or Wiener-Kolmogorov prediction, 
is a widely employed probabilistic interpolation method, 
especially well-suited for noisy and irregularly distributed data.
Originating from early applications in geostatistics, today Gaussian Processes (GPs) have a wide range of applications, primarily in modeling and classification problems; see 
Ref.~\cite{kanagawa2018gaussian} for a review and Ref.~\cite{rasmussen2006gaussian} for a textbook introduction.
Because no explicit functional basis is required, GPR offers a very general approach to interpolating noisy data.
Instead, the characteristics of the underlying function are defined implicitly through the so-called kernel of the GP, making GPR a very flexible non-parametric model when considering interpolations of data with an unknown functional basis.
Here, we will summarize the main concepts and features of the inversion procedure with GPs, 
while we refer to \Cref{app:GPR} for the technical discussion and Refs.~\cite{valentine2020gaussian,Horak:2021syv} for additional details on the method.

In essence, GPs can be viewed as a way of defining a Gaussian distribution over functions, with some prior mean and covariance.
When describing some function as a GP, we can then make prediction about function values by conditioning the GP on the available data.
This is a standard procedure in Bayesian statistics and leads to the posterior predictive distribution.
Due to the probability density being inherently Gaussian, this posterior probability can be computed analytically, and the resulting posterior is again a multivariate normal distribution with a posterior mean and covariance, cf.~\Cref{app:GPR}.
This predictive distribution is a distribution of all possible functions fitting the given data and we use the mean of this distribution in order to make predictions.

The important property that makes GPs practicable for the inversion of integral transformations in particular, is the fact that linear transformations preserve Gaussian statistics.
This means that we can condition the predictive distribution of the GP not only on direct data, but also, in the case of the spectral function, on the correlator data, since the spectral function is connected to the correlator by a linear transformation, cf. \Cref{eq:KL}. 
The resulting predictive distribution is then a distribution over the family of functions which fulfill the integral constraint given by the correlator data. 
This method has been proposed in Ref.~\cite{valentine2020gaussian} and applied to a number of numerical reconstructions in the context of QCD, see e.g.\,Refs.~\cite{Horak:2021syv,Pawlowski:2022zhh,Horak:2023xfb}.

GPR has two main appeals, when considering inverse problems.
Firstly, we do not choose a finite basis for the interpolation.
Instead, we generally focus on GP models that perform the interpolation in an infinite dimensional functional basis.
This means that we can model every continuous function and do not have to assume a prior functional basis.
Secondly, the inclusion of additional data, either direct knowledge of the spectral function or other linearly connected constraints such as the sum rule, e.g.\,\Cref{eq:sumrule}, is straightforward and can further restrict the range of possible functions.
Otherwise, when only considering the correlator data, the uncertainty in the spectral function prediction can be rather large, as expected from an ill-conditioned inversion.

As mentioned before, the GP prior covariance is fully characterized by the so-called GP-kernel.
The GP-kernel constitutes the prior assumption about the function that is interpolated.
Typically, these assumptions encompass very general features, such as continuity and a general length scale.
A popular choice is the radial basis function (RBF) kernel \Cref{eq:RBF}.
As a so-called universal GP-kernel, the associated GP is a universal approximator, meaning that this choice does not restrict the possible spectral functions \cite{steinwart2002, micchelli06a}.
However, when having additional information about the functional structure of the spectral function, such as UV asymptotics, the GP-kernel can be extended to incorporate this information in certain regions, as described in \Cref{app:GPR}.

\subsubsection{GPR Reconstruction Details}\label{sec:GPRecon}
%
\begin{figure*}
    \centering
    \includegraphics[width=8cm]{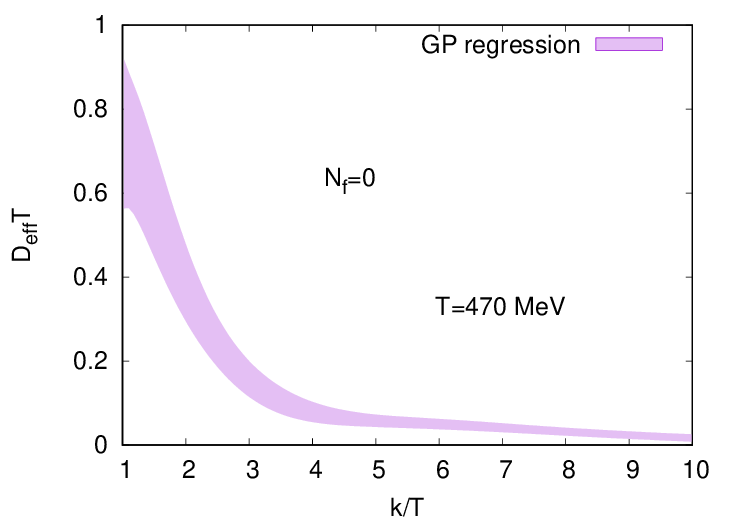}
    \includegraphics[width=8cm]{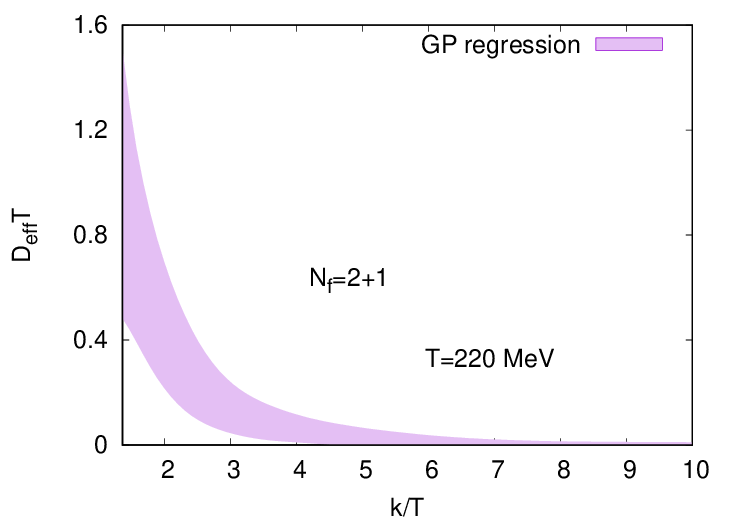}
	\caption{
	Effective diffusion coefficent $D_\text{eff}$ obtained, as in \Cref{eq:Deff},
	from the GPR reconstruction of $\rho_H$ for quenched on the left and for full QCD on the right. The error band includes both the error propagation of the lattice data and the systematic error from the GP model.}
    \label{fig:DEFF_GP}
\end{figure*}

Since we have data for several spatial momenta, the reconstruction of the spectral function is not only performed in the $\omega$, but also simultaneously in the $k$-direction.
This ensures continuity in both directions and consequently increases the stability of the reconstruction.
However, apart from continuity, we do not assume any additional structure of the spectral function in the momentum direction.
For the reconstruction, we include the lattice data for all the available spatial momenta as well as the sum rule \Cref{eq:sumrule}.

As a very common GP-kernel choice, the RBF kernel \Cref{eq:RBF} is used.
In general, we observe that choosing different smooth, stationary and universal GP-kernels have very little impact on the reconstruction.
In order to include the known UV asymptotics, $\rho_H^\tinytext{UV}(\omega) \propto 1 / \omega^4$, the GP-kernel is modified in the UV regime to restrict the functional basis to the asymptotic behavior resulting in \Cref{eq:uvkernel}.
The transition from the RBF kernel to the UV kernel is controlled by smooth step functions, for more details on the exact form of the resulting GP-kernel see \Cref{app:GPR}.
Combining the RBF kernel in two dimension with the asymptotic kernel results in a total number of five GP-kernel parameters: $l_\omega,\,l_k,\,\sigma,\,l_\tinytext{UV},\,\nu$.
The lenghtscale in $k$ and $\omega$ direction is controlled by $l_k$ and $l_\omega$ respectively, while $\sigma$ gives a prior estimate on the variance of the GP.
The position of the transition from the RBF kernel towards the UV kernel is controlled by 
$\omega_0 =\sqrt{k^2 + \nu (\pi T)^2}$, where $\nu$ is varied, 
in analogy to the polynomial fit ansatz, and $l_\tinytext{UV}$ is the smoothness of the transition between the two kernels.

These parameters are optimized by minimizing the associated negative log-likelihood (NLL) \Cref{eq:NLL}.
Since this optimization is performed in a five dimensional parameter space and the parameters are not fully independent, this optimization generally does not converge consistently.
Additionally, open directions in the parameter space, e.g.\,towards vanishing length scales are possible.
This can be viewed as a manifestation of the inherent ill-conditioning of this inverse problem.
Therefore, and in order to capture the systematic error from the introduction of the UV kernel, we perform a Metropolis-Hastings sampling of the NLL in the parameter space, see \Cref{app:GP_params} for details.
While the error on the lattice data is propagated through the GP and is given by the covariance of the posterior predictive distribution of the GP, the systematic error is estimated by the uncertainty in the parameters while minimizing the associated NLL.
This systematic error is mostly the result of introducing an explicit functional basis in the UV, while a change in the RBF kernel parameters results in a much smaller deviation in the spectral function. 
The final error estimate on the spectral function from the GP reconstruction is given by a combination of the statistical and systematic error.

The resulting spectral functions corresponding to the quenched and full QCD lattice correlator are presented in \Cref{gp_spectral}.
In order to compare with the reconstructions from the fits and the BG method, we have evaluated the spectral function at the spatial momenta $k$, where the lattice data is available.
In general however, the GP gives a statistical estimate for all momenta in between the available data.
This gives a continuous estimate on the thermal photon rate not only at the given spatial momentum values, but also in between, see \Cref{fig:DEFF_GP}.
The validity of the interpolation between data points has been additionally confirmed by reconstructing the effective diffusion coefficient while disregarding lattice data for a single spatial momentum.  
This gives, as expected, very similar results with an increased error around this momentum value.
Although the interpolation in momentum-direction give reliable estimates, the extrapolation towards smaller or higher momenta comes with large uncertainties.
Ultimately, when extrapolating towards $k=0$, we recover the GP prior as the current assumptions about the spectral function do not allow for a systematic extrapolation.
The resulting values for $D_\text{eff}T$ from GP reconstruction are given in
\Cref{Tab:Deff quenched} for quenched and in 
\Cref{Tab:Deff 3 flavour} for full QCD.

%% file: Deff.tex
\begin{figure*}
    \centering
    \includegraphics[width=8cm]{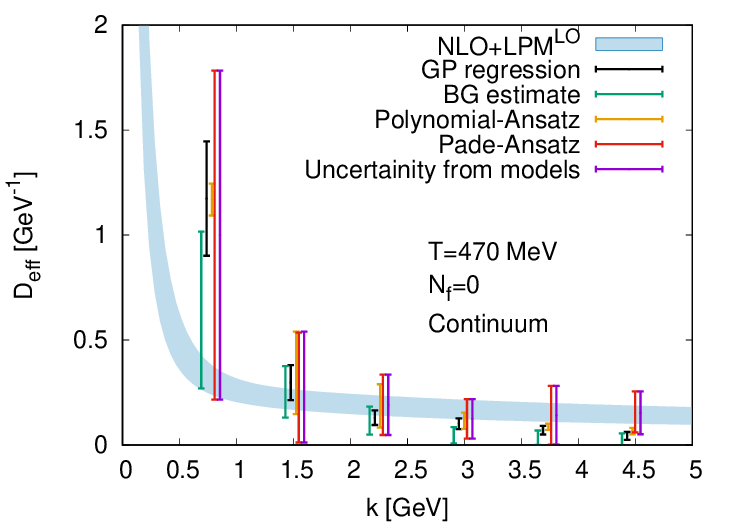}
    \includegraphics[width=8cm]{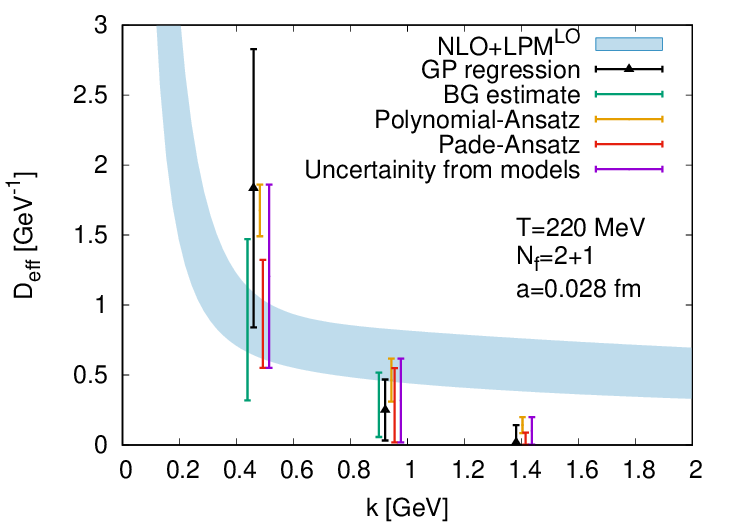}
	\caption{
	Effective diffusion coefficient from all spectral reconstruction methods studied in this paper. 
	The $N_f=0$ case appears on the left panel,  $N_f=2+1$ on the right.
	Here we have converted the results in physical units. For better visibility, error bars at the same momentum value are slightly shifted on the horizontal axis. 
	}
    \label{DEFF}
\end{figure*}

In the previous sections we discussed various models and methods for the spectral reconstruction from lattice data. Now, we compare the photon production rate obtained from these different methods.

In \Cref{DEFF}, we show the effective diffusion coefficient as a function of momentum for all the reconstruction methods we performed. We observe that the uncertainty in the estimation of $D_{\text{eff}}$ is larger at small momenta. This is expected because at small momenta the lattice correlators are flat compared to the larger momenta, corresponding to a narrow peak in $\rho_H/\omega$ at $\omega=0$ for small momenta. On the other hand, at larger momenta, $D_{\text{eff}}$ shows better agreement, as different methods provide closer estimates, while the relative error on the reconstructions remains approximately constant. 

On the same plot, we also display the corresponding perturbative estimate of $D_{\text{eff}}(k)$. The convergence of the perturbative estimate can be checked by varying the renormalization scale $\mu$.
Here, we estimate the error band by varying the scale from\footnote{
The smaller value of $\mu$ corresponds to an upper estimate of $D_{\rm eff}\,$, due to the larger QCD coupling being probed. In our case, the extreme choice of $\mu=0.5\,\mu_\tinytext{opt}$ leads to $\alpha_s \simeq 0.5$ which is where most of the sensitivity to scale variation originates.
} $\mu=0.5\,\mu_\tinytext{opt}$ to $\mu = 2 \,\mu_\tinytext{opt}$ (where the `optimal' scale is defined in \Cref{sec:comp_pt}). 
The error band on the perturbative estimate is slightly smaller for quenched QCD at $T=\SI{470}{\mega\eV}$ compared to full QCD at $T=\SI{220}{\mega\eV}$, which may indicate that the perturbative framework 
is better behaved in this case.

In the case of quenched QCD, our non-perturbative estimate of $D_{\text{eff}}$ from the Padé-like ansatz is consistent with the perturbative estimate within the error bar. 
However, at the lowest momentum, the Polynomial-ansatz and GP regression favor a higher photon rate compared to the perturbative estimate, whereas at large momentum, the polynomial-ansatz, BG, and GP estimates give a lower photon rate than the perturbative estimate. 
When it comes to full QCD, we observe in all the non-perturbative estimates a significantly more rapid decrease in the photon production rate compared to the perturbative estimate . 

This aligns with the fact that, in units of the respective transition temperature, quenched QCD is at approximately $25\%$ higher temperature than full QCD. As a result, one would expect that the quenched QCD result will be closer to the perturbative result than the full QCD case. 

It is important to note that the perturbative calculation of the spectral function on the light cone relies on a coupling controlled entirely by the temperature. 
Consequently, one should expect the non-perturbative estimate to approach the perturbative estimate only at high temperatures and for momenta $k\sim T$. 

We combined the $D_{\text{eff}}$ from different models to obtain an overall error bar arising from model estimates. This is done by spanning the highest and lowest value of $D_{\text{eff}}$ from models for a given momentum $k$. 

Qualitatively, we can compare our dynamical results with other lattice studies of $D_{\text{eff}}$. Our results are in qualitative agreement with the study  in Ref.~\cite{Ce:2020tmx} at a temperature of $\SI{254} {\mega\eV}$ (1.2 $T_{pc}$). In both studies, at the lowest momentum ($k=\SI{0.4}{\giga\eV}$), $D_{\text{eff}}$ ranges from ${\SI{0.5}{\giga\eV}^{-1}}$ to ${\SI{2.7}{\giga\eV}^{-1}}$. At the highest available momentum, the results are consistent with zero within the error bars. However, the error on our $D_{\text{eff}}$ is smaller for the highest available momentum compared to Ref.~\cite{Ce:2020tmx}. This could be due to the different systematics involved in the spectral reconstruction methods, as well as the fact that the results in Ref.~\cite{Ce:2020tmx} are from continuum-extrapolated 2-flavor QCD correlators, whereas the dynamical results in this paper are from (2+1)-flavor QCD at a fixed lattice spacing.

In Ref~\cite{Ce:2022fot}, the photon production rate is estimated from the transverse channel of the correlator. In this case, the results are consistent at smaller momentum. However, $D_{\text{eff}}$
  estimated here shows a smaller value at the highest momentum, which could again be due to different systematics involved, similar to the reason mentioned above.

\begin{table*}
\centering
\begin{tabular}{
    @{\hspace{0.5em}}c@{\hspace{1.5em}}c@{\hspace{1.5em}}
    c@{\hspace{1.5em}}
    c@{\hspace{1.5em}}
    c@{\hspace{.5em}}
  } 
\hline \\[-1.5mm]
$k/T$ & 
  polynomial & Pad\'e  & Backus-Gilbert & GP regression
  \\[1.5mm]
\hline
 \\[-2.5mm]
$1.57$ & $0.549 \pm 0.035$ & $0.469 \pm 0.368$ & $0.302 \pm 0.175$ & $0.552 \pm 0.129$ \\[0.5mm]
$3.14$ & $0.161 \pm 0.092$ & $0.129 \pm 0.123$ & $0.119 \pm 0.057$ & $0.140 \pm 0.039$ \\[0.5mm]
$4.71$ & $0.087 \pm 0.048$ & $0.089 \pm 0.067$ & $0.054 \pm 0.031$ & $0.061 \pm 0.016$ \\[0.5mm]
$6.28$ & $0.054 \pm 0.018$ & $0.058 \pm 0.044$ & $0.022 \pm 0.018$ & $0.048 \pm 0.012$ \\[0.5mm]
$7.85$ & $0.041 \pm 0.007$ & $0.066 \pm 0.066$ & $0.016 \pm 0.016$ & $0.033 \pm 0.010$ \\[0.5mm]
$9.42$ & $0.031 \pm 0.007$ & $0.074 \pm 0.045$ & $0.013 \pm 0.013$ & $0.020 \pm 0.009$ \\[0.5mm]
\hline
\end{tabular}
  \caption{
  Extracted values of $D_{\text{eff}}T$ from
  different methods for $N_f=0\,$.
  }
  \label{Tab:Deff quenched}
\end{table*}

\begin{table*}
  \centering
\begin{tabular}{
    @{\hspace{0.5em}}c@{\hspace{1.5em}}c@{\hspace{1.5em}}
    c@{\hspace{1.5em}}
    c@{\hspace{1.5em}}
    c@{\hspace{.5em}}
  } 
\hline \\[-1.5mm]
$k/T$ & 
  polynomial & Pad\'e  & Backus-Gilbert & GP regression
  \\[1.5mm]
\hline
 \\[-2.5mm]
$2.09$ & $0.369 \pm 0.040$ & $0.206 \pm 0.085$ & $0.197 \pm 0.127$ & $0.404 \pm 0.219$ \\[0.5mm]
$4.19$ & $0.102 \pm 0.034$ & $0.062 \pm 0.058$ & $0.063 \pm 0.047$ & $0.055 \pm 0.048$ \\[0.5mm]
$6.28$ & $0.031 \pm 0.013$ & $0.010 \pm 0.010$ & $-0.011 \pm0.047 $ & $0.004 \pm 0.027$ \\[0.5mm]
\hline
\end{tabular}
  \caption{
  Extracted values of $D_{\text{eff}}T$ from
  different methods for $N_f=2+1\,$.
  }
  \label{Tab:Deff 3 flavour}
\end{table*}

%% file: conclusion.tex
We estimated the thermal photon production rate from the Euclidean lattice correlators in quenched QCD ($T=\SI{470}{\mega\eV}$) and full QCD ($T=\SI{220}{\mega\eV}$) with an unphysical pion mass of $\SI{320}{\mega\eV}$. Motivated by Ref.~\cite{Ce:2020tmx}, we have utilized the T-L correlator for calculating the photon rate. This choice offers several advantages over the vector correlator, primarily due to the absence of a significant UV component in the T-L correlator. The spectral function of this correlator satisfies a sum rule along with a known asymptotic form in the UV. 
In the quenched case this T-L correlator has been extrapolated to the continuum whereas in full QCD the results are from a single lattice spacing.  
We studied non-perturbative effects in this correlator by comparing it with its perturbative estimate, shown in \Cref{corr_per_latt}. \Cref{per_sp} displays the perturbative spectrum, which was calculated at NLO away from the light cone, while leading-order LPM resummation has been done near the light cone~\cite{Jackson:2019yao}. 

We observe that in quenched QCD, the non-perturbative effects are much more pronounced at lower momenta compared to higher momenta. This is evident as the lattice correlators are significantly flatter in contrast to the perturbative correlator. As we move towards larger momenta, the lattice data also begins to exhibit curvature similar to the perturbative correlator. However, it appears that non-perturbative effects persist even at the highest available momentum we have.

In the case of full QCD, the lattice data displays a substantially flatter behavior (as a function of $\tau$) compared to the perturbative correlator, indicating significant non-perturbative modification.

To obtain photon rate, we performed the spectral reconstruction using the lattice correlators we determined. Since the inverse problem is an ill-conditioned problem, we used multiple models and methods for the spectral reconstruction. In all models, we constrained the spectral function according to the sum rule and incorporated the known asymptotic form. The first method we used is a simple polynomial ansatz of the spectral function in the IR region, which is based on the smoothness of the spectral function across the light cone. We smoothly connected the IR part with a UV part of the spectral function consistent with the OPE expansion, as given in \Cref{poly-ansatz} . This spectral function also satisfies the sum rule. Using this form we performed the fitting with the lattice data with respect to the free parameters of the model. The resulting spectral function is plotted in \Cref{Quenched_polynomial_sp} along with the perturbative spectral function. We see that at small momentum, the spectral function has a larger dependence on the matching point $\omega_0$ compared to the case of large momentum. As the correlator is very flat at small momentum, it signifies a very narrow peak at the origin, which is difficult to determine.\footnote{The uncertainty with respect to the variation of $\omega_0$ is taken to be the systematic uncertainty on our result.} Here we also observed that except the smallest momentum, for the quenched case the resulting non-perturbative spectral function is much closer to the perturbative spectral functions compared to the full QCD case. This is expected because effectively the temperature for the quenched case is larger than the full QCD case. As a second model we used a Pad\'e-like ansatz of the spectral function as put forward in \cite{Ce:2020tmx}. A sample spectral function is shown in \Cref{pade_sp}. 

We also used the well-known BG method, to obtain a smeared spectral function, which is a convolution of the actual spectral function with a resolution function. The additional prior function we used is given in \Cref{priorbg}, whereby it contains the known asymptotic behavior of the spectral function. This method has the stability parameter $\lambda$ and the parameter $\nu$ from the prior function. We found that the photon production rate is stable w.r.t. the variation of  $\lambda$ and $\nu$ for all the momenta. For the lowest available momentum we found that although the result is stable with respect to $\lambda$, it shows some dependence on $\nu$, while the photon production rate nevertheless saturates at large $\nu$. This variation with respect to $\nu$ has been taken as a systematic error on the photon production rate.

As a third method we applied Gaussian Process regression in order to obtain a prediction for the spectral function. Instead of giving a singular prediction, this method returns a distribution over functions that agree with the given correlator data. The T-L correlator data is reconstructed continuously in both $\omega$ and $k$ directions, giving a continuous prediction for the effective diffusion coefficient, see \Cref{fig:DEFF_GP}. By considering the UV asymptotic behavior from the OPE and the sum rule, we can reduce the variance on the GP posterior in order to make meaningful predictions.  

The photon production rate in terms of $D_{\text{eff}}$ is plotted in \Cref{DEFF}. The error bar from the models have been combined to provide a single error bar. 
As momentum increases, the results from different methods begin to converge within their systematic error bars. In the quenched case, we observe that the lattice estimate of the photon production rate approaches the perturbative estimate. In the full QCD case, the photon production rate becomes smaller than the perturbative estimate and becomes consistent with zero at a momentum of approximately $\SIrange[range-phrase = \text{\,...\,}]{0.8}{1.6}{\giga\eV}$.

There are many ways one can proceed for future studies. Currently, the full QCD results are obtained at finite lattice spacing. For larger momentum, the cutoff effects could be significant, and a continuum extrapolation is needed. The full QCD results in this paper are based on unphysical pion mass configurations. To make predictions aligned with experimental results, it is important to calculate the photon rate using physical pion mass gauge configurations. Although it is expected that at high temperatures, the effect of quark mass on the photon rate should be small. It is also important to calculate the photon rate for different temperatures, which will be relevant as input to hydrodynamics.

\section{Acknowledgments}

We would like to thank Luis Altenkort and Hai-Tao Shu for their work on integrating the mesonic correlator into the QUDA code and generating gauge field configurations using SIMULATeQCD~\cite{HotQCD:2023ghu}. D.~B. would like to thank Guy D. Moore for various discussions on the photon production rate. D.~B. would also like to appreciate the discussion with Harvey B. Meyer during the Lattice 2022 conference for the discussion on sum rules and asymptotic behavior of the spectral function. S.~A., D.~B., O.~K., J.~T., T.~U. and N.~W. are supported by the Deutsche Forschungsgemeinschaft (DFG, German Research Foundation) - Project number 315477589-TRR 211. N.~W. acknowledges support by the State of Hesse within the Research Cluster ELEMENTS (Project ID 500/10.006). For the computational work We used the Bielefeld GPU cluster and the JUWELS supercomputer.
The authors gratefully acknowledge the Gauss Centre for Supercomputing e.V. for funding this project by providing computing time through the John von Neumann Institute for Computing (NIC) on the GCS Supercomputer JUWELS at Jülich Supercomputing Centre (JSC).
G.~J. was funded by the U.S. Department of Energy (DOE), under grant No.~DE-FG02-00ER41132, and now by the ANR under grant No.~ANR-22-CE31-0018 (AUTOTHERM). A.~F. acknowledges support by the National Science and Technology Council of Taiwan under grant 111-2112-M-A49-018-MY2.

%% file: app.tex
\section{Mock-analysis of perturbative data with Polynomial ansatz}\label{app:mock_poly}
%
\begin{figure*}
    \centering
    \includegraphics[width=8cm]{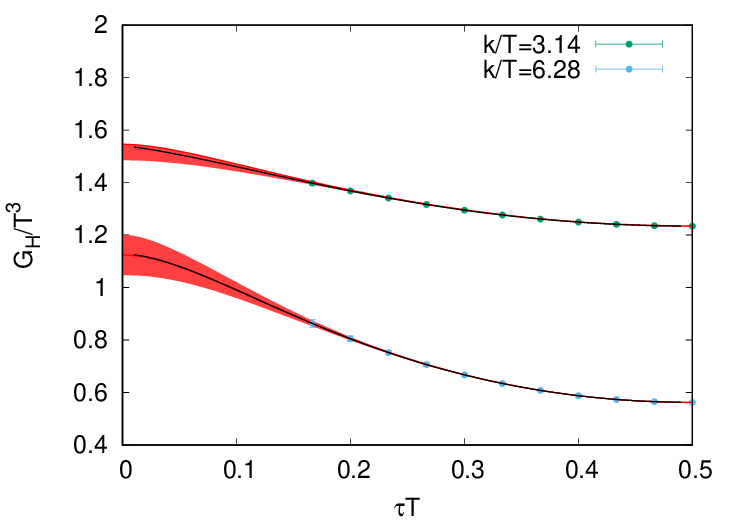}
    \includegraphics[width=8cm]{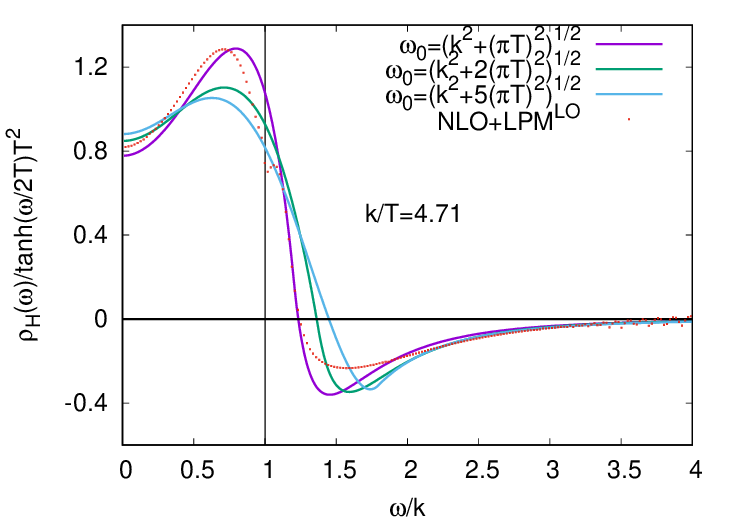}
	\caption{
	Mock analysis of the data generated from the perturbative correlator. In the left panel, the red band represents the fitting of the data points with the polynomial ansatz, while the black continuous line represents the \NLPM\ perturbative correlator. In the right panel, the reconstructed spectral function is displayed along with the exact \NLPM\ spectral function.}
    \label{mock}
\end{figure*}
In this section, we conduct a mock analysis to demonstrate the effectiveness of the polynomial ansatz fitting. To generate the mock data set, we use the perturbative \NLPM\ correlator, which is obtained from the known perturbative spectral function $N_f=0$ case. In the mock data, we include an error bar equivalent to the corresponding lattice correlator. 
We used 11 equally spaced data points in the interval from $\tau T \simeq 0.17$ to $\tau T = 0.5\,$, 
matching the number of continuum-extrapolated lattice data points for $N_f=0$.

We perform the fitting with the polynomial ansatz for the spectral function and obtain the fit, which is shown in the left panel of \Cref{mock} for the momentum $k/T=3.14$ and $k/T=6.28$. The black line corresponds to the exact correlator, while the red band corresponds to the fitting of the spectral function for $\omega_0=\sqrt{k^2+(\pi T)^2}$.

The right panel of \Cref{mock} shows the resulting spectral function for various values of $\omega_0\,$ at $k/T=4.71$. 
We can see that this ansatz is indeed able to approximately predict the perturbative spectral function within the range of
$\omega_0=\sqrt{k^2+\nu(\pi T)^2}$ where $\nu = 1\,...\,5\,$.

\section{\texorpdfstring{$D_\text{eff}$ from finite lattice spacing for $N_f=0$}{Deff from finite lattice spacing for Nf=0}}\label{app:finite_latt_spacing}

\begin{figure*}
    \centering
    \includegraphics[width=8cm]{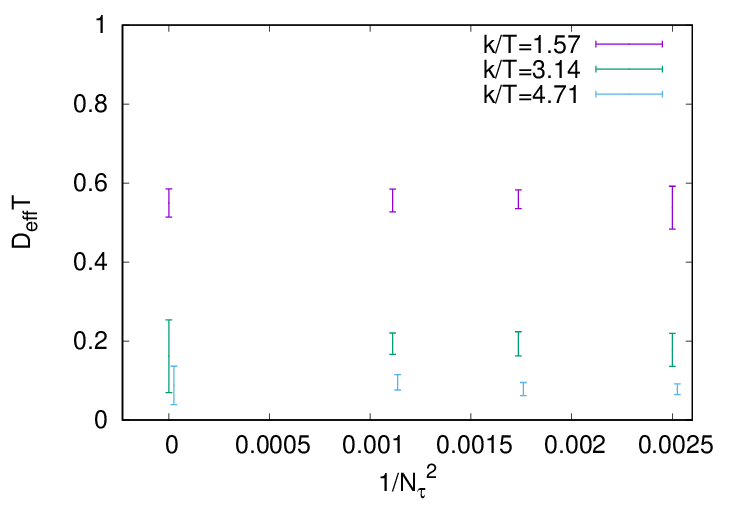}
    \includegraphics[width=8cm]{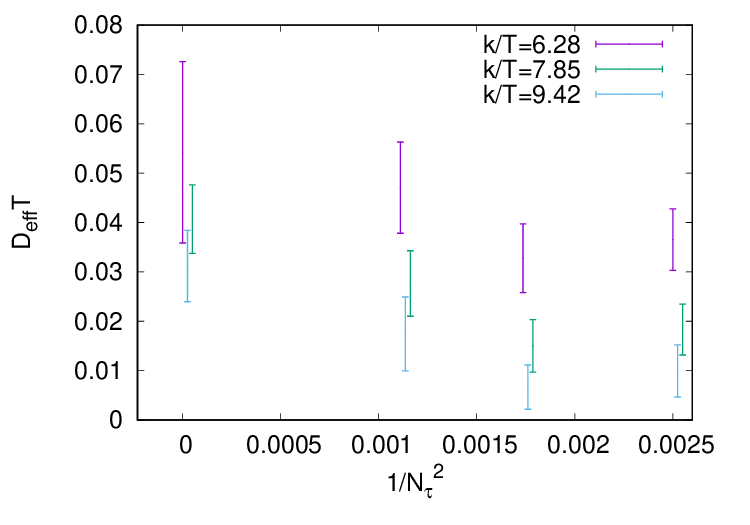}
	\caption{
	Effective diffusion coefficient, $D_\text{eff}$ obtained from the polynomial estimate for different lattice spacing along with continuum extrapolated value.
	Here $N_f = 0\,$.}
    \label{fig:sp_cut_off}
\end{figure*}

\begin{figure*}
    \centering
    \includegraphics[width=8cm]{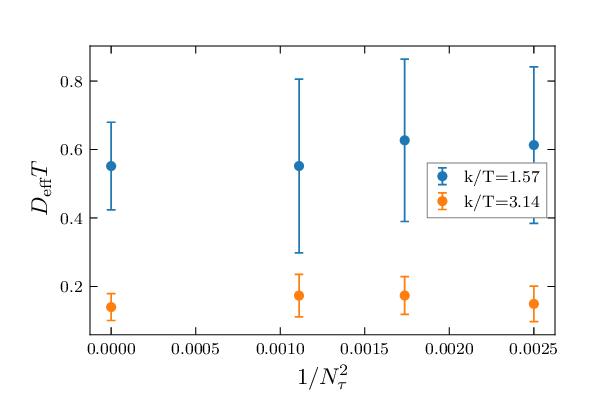}
    \includegraphics[width=8cm]{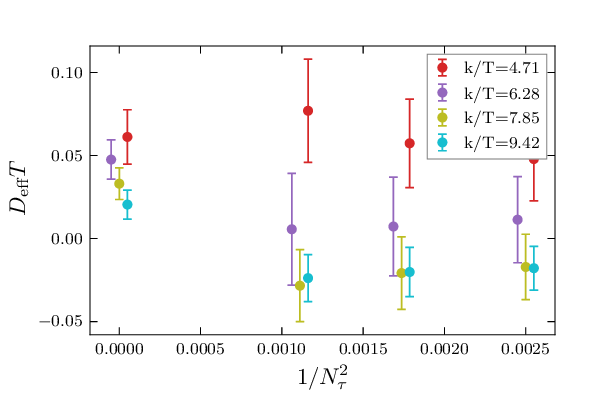}
	\caption{
	Effective diffusion coefficient, $D_\text{eff}$ obtained from the GP regression model for different lattice spacing along with continuum extrapolated value.
	Here $N_f = 0\,$.
	}
    \label{fig:cutoff_GP}
\end{figure*}

In this section, we will compare the estimate of the effective diffusion coefficient from different lattice spacings in order to estimate the cutoff effects on the reconstruction.
This will be illustrated for the polynomial ansatz of the spectral function and the GPR reconstruction.
A comparison of the different $D_\text{eff}$ values from the polynomial ansatz can be found in \Cref{fig:sp_cut_off}, while the same comparison for the GPR reconstruction can be found in \Cref{fig:cutoff_GP}.
We observe that for both methods the results for small spatial momenta agree remarkably well, as different values remains consistent in the margin of error.
For the polynomial ansatz and spatial momenta $k/T>4.71$ we observe mild cutoff effects on $D_\text{eff}$, while in the GP reconstruction, the error generally grows for higher momenta.
This outcome aligns with our expectations, considering that cutoff effects are already minimal in the correlator.
%
\section{Inversion with GPR}\label{app:GPR}
%
\begin{figure*}
    \centering
    \includegraphics[width=16cm]{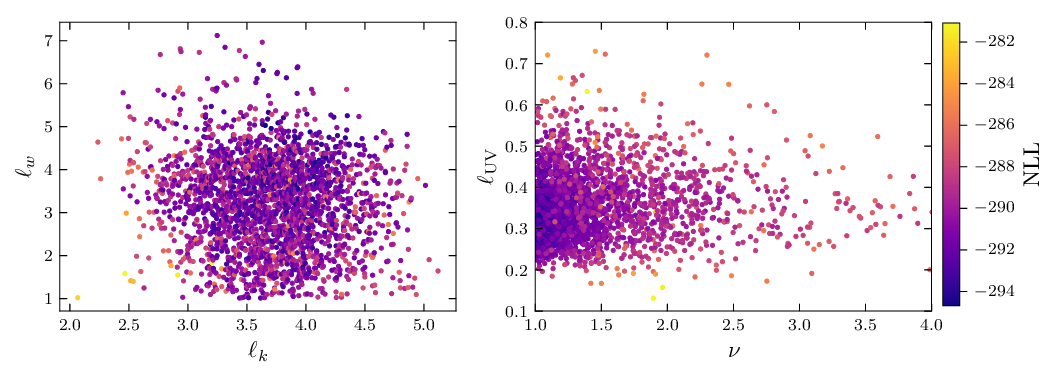}
	\caption{Metropolis-Hastings sampled GP kernel parameters in the negative log-likelihood plane.}
    \label{fig:GP_param_scan}
\end{figure*}
Here, we will give a short introduction to the reconstruction of spectral functions with GPR.
For additional details and a more general discussion, we refer to Refs.~\cite{valentine2020gaussian,Horak:2021syv} and to Ref.~\cite{rasmussen2006gaussian} for a introduction to GPs in general.
Note that in the following part, we closely follow the introduction to GP reconstruction found in Ref.~\cite{Horak:2023xfb} and modify the notation to match the current reconstruction in order to ensure a self-contained and reproducible presentation. 

To start, we define a GP prior for the spectral function that is reconstructed.
This assumes some Gaussian distribution over different realizations of the spectral function.
The GP prior is generally written as 
\begin{align}
    \rho(\omega) \sim \mathcal{GP}\big(\mu(\omega), C(\omega, \omega)\big)
\,,
\end{align}
where $\mu$ and $C$ denote the mean and the covariance of the GP prior.
The prior mean can be generally set to zero, as its contribution can be fully absorbed into the covariance.

For the reconstruction, the GP prior is then conditioned on the available data, e.g. observations of the propagator $G_{H,i}$ at discrete Euclidean times $\tau_i \equiv [\boldsymbol{\tau}]_i$.
This means that we compute the conditional posterior distribution of the GP, which has a closed analytic form and is given by \cite{Horak:2021syv} 
\begin{align}
    \rho(\omega)|G(\boldsymbol{\tau}) \sim &\, \mathcal{GP}\Bigl(\boldsymbol{w}^\intercal(\omega)(\boldsymbol{W} + \sigma_n^2 1)^{-1}G_H(\boldsymbol{\tau}), \nonumber \\[1ex]
    & \hspace{-.5cm}C(\omega,\omega) - \boldsymbol{w}^\intercal(\omega)(\boldsymbol{W} + \sigma_n^2 1)^{-1}\boldsymbol{w}(\omega)\Bigr)
\,,
\end{align}
where
\begin{align}\label{eq:posterior}
    &[\boldsymbol{w}]_i(\omega) = \int \text{d}\omega'\, K(\tau_i, \omega') C(\omega', \omega)\,,\nonumber \\[1ex]
    &[\boldsymbol{W}]_{ij} = \int\text{d}\omega'\text{d}\omega'' \, K(\tau_i, \omega') K(\tau_j, \omega'') C(\omega', \omega'')
\,.
\end{align}
and $K(\tau, \omega) \equiv \frac{\cosh(\omega(1 / 2T - \tau))}{\pi \sinh(\omega/2T)}$.
\Cref{eq:posterior} is a standard result in multivariate statistics and is essentially equivalent to GPR with direct observation only with the addition of the integral transformation, that is supposed to be inverted.
This posterior encodes the knowledge of the spectral function under the constraint of the correlator observation and directly accounts for the error estimations on the correlator, here denoted as $\sigma_n$.

The prior covariance of the GP is characterized by the so-called kernel.
This kernel is generally a function with a small number of hyperparameters and fully characterizes the GP.
The kernel controls very general features of the interpolation, such as differentiability of the function, a characteristic lengthscale of the underlying function or an estimate on the variance of the interpolation.
A commonly used kernel choice is the radial basis function (RBF) or Gaussian kernel, defined as
\begin{align} \label{eq:RBF}
    C_{\tinytext{RBF}}(\omega, \omega') = \sigma^2 \exp\left(-\frac{(\omega-\omega')^2}{2l^2}\right)
    \,,
\end{align}
where $\sigma$ controls the overall magnitude and $l$ is a generic lengthscale within the GP.
The RBF kernel falls under the class of so-called universal kernels \cite{steinwart2002, micchelli06a}.
Such kernels have the universal approximation property; any continuous function can be approximated by the corresponding GP.

The kernel can be extended in order to capture knowledge about the asymptotics of the spectral function, as proposed in Ref.~\cite{Horak:2023xfb}.
The main idea is to reduce the basis function of the GP kernel in a certain region to just represent the known asymptotics of the spectral function.
For the known UV asymptotics $\rho_H^\tinytext{UV}(\omega) \propto 1/\omega^4$ of the spectral function, this can be achieved by taking the asymptotic kernel
\begin{align} \label{eq:uvkernel}
    C_\tinytext{UV}(\omega, \omega') \; = \; 
    \big(\omega \, \omega^\prime \big)^{-4}
    \,.
\end{align}
In order to have a smooth transition between the universal and the asymptotic kernel, the transition is controlled by a soft step function
\begin{align}\label{eq:step}
    \theta^\pm(\omega; l_\tinytext{UV}, \omega_0) = \frac{1}{1 + \exp(\pm(\omega - \omega_0)) / l_\tinytext{UV}}
    \,,
\end{align}
where $\omega_0$ is the position and $l_\tinytext{UV}$ controls the steepnes of the transition from the universal to the UV kernel.
This results in the full kernel
\begin{align} \label{eq:fullkernel}
    C(\omega, \omega') &= \theta^-(\omega)\theta^-(\omega') C_\tinytext{RBF}(\omega, \omega')\nonumber \\[1ex]
    &+ \theta^+(\omega)\theta^+(\omega')C_\tinytext{UV}(\omega, \omega')
    \,.
\end{align}
The parameters of the soft step function, i.e.\,the position and the lengthscale of the transition, and of the RBF kernel are subject to likelihood optimization, described in the following.  

\subsection{GPR parameter optimization}\label{app:GP_params}

The GP kernel parameter optimization is a pivotal part of the reconstruction, since the kernel parameters control the implicit structure of the reconstructed function.
In order to optimize these parameters, the negative log-likelihood (NLL) of the associated GP is minimized.
The NLL of a general GP, with matrices as defined in \Cref{eq:posterior}, is given by
\begin{align}\label{eq:NLL}
	-\log p(G_H(\boldsymbol{\tau})|\boldsymbol{\sigma}) = \frac{1}{2} G_H(\boldsymbol{\tau})^\intercal \left(\boldsymbol{W_\sigma} + \sigma_n^2 1 \right)^{-1} G_H(\boldsymbol{\tau})\nonumber \\[1ex] 
	+\frac{1}{2}\log \det(\boldsymbol{W_\sigma} + \sigma_n^2 1) + \frac{N}{2} \log 2\pi
    \,,
\end{align}
where the dependence of the kernel on the hyperparameters is indicated by $\boldsymbol{\sigma}$.
These parameters are optimized by performing a Metropolis-Hastings scan of the NLL in the five dimensional parameters space spanned by $l_\omega,\,l_k,\,\sigma,\,l_\tinytext{UV},\,\nu$, presented in \Cref{sec:GPRecon}.
$\sigma$ is the overall magnitude of the kernel, while $l_\omega$ and $l_k$ are the RBF lengthscales in $\omega$ and $k$ direction respectively.
$l_\tinytext{UV}$ is the steepness of the asymptotic transition, while $\nu$ controls the position of the transition in \Cref{eq:step} with $\omega_0 = \sqrt{k^2 + \nu (\pi T)^2}$, in analogy to the polynomial fit ansatz.

When optimizing the parameters for the one dimensional GP, e.g.\,for one fixed spatial momentum, we notice that the overall magnitude $\sigma$ of the kernel decreases with $1/k^2$, while the other parameters fluctuate, but do not have a qualitatively different behavior.
We therefore rescale the magnitude parameter for the two-dimensional reconstruction as $\sigma\rightarrow\sigma/k^2$.
In \Cref{fig:GP_param_scan}, the parameters scan of the NLL is shown.
We find a mild dependence of the spectral function on the lengthscales and the magnitude of the RBF kernel, e.g.\,$l_\omega,\,l_k,\,\sigma$, while the dependence on the asymptotic parameters, as expected, introduces the majority of the systematic error.
The systematic error, captured by varying these hyperparameters has similar values to the error propagation on the error of the underlying lattice data.
The parameter space is limited to values of $\nu\geq 1$, since the perturbative behavior does not reach this far into the IR.
However, such solutions are still represented by the universal RBF kernel, we merely do not want to  prefer the min this region of the spectral function.
Additionally, the lengthscale of the transition is also restricted from below, $l_{\tinytext{UV}}\geq0.1$, in order to avoid edges in the reconstructed spectral function.
These open directions in the parameter space can be attributed to inversion being ill-conditioned.
The error on the reconstructed spectral function is finally computed as a combination of the intrinsic error propagation of the GP, which takes the lattice data error into account, and the variation of the spectral function with the parameters given by the NLL scan.
